\documentclass[twocolumn,preprintnumbers,amsmath,amssymb]{revtex4}
\usepackage{epsfig,graphicx}
\usepackage{dcolumn}
\usepackage{bm}
\begin{document}
\draft
\title{Pair dynamics in a glass forming binary mixture: Simulations and theory}

\vspace{1cm}
\author{Rajesh K. Murarka and Biman Bagchi\footnote[1]{For correspondence: bbagchi@sscu.iisc.ernet.in}}
\address{Solid State and Structural Chemistry Unit,\\
Indian Institute of Science,\\
Bangalore, India 560 012.}

\begin{abstract}

We have carried out molecular dynamics simulations to understand the dynamics of a {\it tagged pair 
of atoms} in a strongly non-ideal glass-forming binary Lennard-Jones mixture. Here atom B is 
smaller than atom A ($\sigma_{BB} = 0.88\sigma_{AA}$, where $\sigma_{AA}$ is the molecular diameter
of the A particles) and the AB interaction is
stronger than that given by Lorentz-Berthelot mixing rule ($\epsilon_{AB} = 1.5\epsilon_{AA}$, where
$\epsilon_{AA}$ is the interaction energy strength between the A particles).
The generalized time-dependent pair distribution function is calculated separately for 
the three pairs (AA, BB and AB). The three pairs are found to 
behave differently. The relative diffusion constants are 
found to vary in the order $D_R^{BB} > D_R^{AB} > D_R^{AA}$, with $D_R^{BB} \simeq 2D_R^{AA}$, showing the importance of the hopping process (B hops much more than A). We introduce a {\it non-Gaussian 
parameter} ($\alpha_2^P (t)$) to monitor the relative motion of a pair of atoms, and evaluate 
it for all the three pairs, with initial separations chosen to be at the first 
peak of the corresponding partial radial distribution functions. At intermediate times, 
significant deviation from the Gaussian behavior of the pair distribution
functions is observed, with different degree for the three pairs. A simple mean-field (MF) model,
proposed originally by Haan [Phys. Rev. A ${\bf 20}$, 2516 (1979)] for one component liquid, is 
applied to the case of binary mixture, and compared with the simulation results. While the MF model 
successfully describe the dynamics of the AA and AB pair, {\it the agreement for the BB pair is less 
satisfactory}. This is attributed to the large scale anharmonic motions of the B particles in a weak 
effective potential. Dynamics of next nearest neighbor pairs are also investigated.
     
\end{abstract}
\maketitle

\section{Introduction}

In dense fluids, there are many interaction-induced phenomena that can be interpreted 
in terms of the dynamics of the pairs of atoms\cite{bloom,litovitz,miller}. 
For example, nuclear overheusser effect
studies the relative motion of the atoms. In addition, an understanding of pair 
dynamics can be of great importance in the studies of rate of various diffusion controlled 
chemical reactions in dense fluids\cite{adelman,hynes}. Both the theoretical 
analysis\cite{bloom,haan,boley,balu1,balu2,posch,kivelson} and 
computer simulation\cite{haan,balu1,balu2,posch,chang} 
studies have been carried out extensively to study the dynamics of a pair of atoms in an 
one component liquid. Surprisingly, however, we are not aware of any explicit study of the 
dynamics of atomic pairs in binary mixtures whose dynamics generally show strong nonmonotonic 
composition dependence\cite{arnab1,rm1}.

The study of the electronic spectroscopy of dilute chromophores ('solutes') in fluids ('solvents') 
is a useful tool for obtaining the information about the structure and dynamics of the solvents 
in the vicinity of the solute. In an attempt to provide a microscopic foundation of the Kubo's 
stochastic theory of the line shape, Skinner and coworkers\cite{skinner} have recently developed 
a molecular theory for the absorption and emission line shapes and ultrafast solvation dynamics 
of a dilute nonpolar solute in nonpolar fluids. Due to the motion of the solvent molecules 
relative to the chromophore, the chromophore's transition frequency generally fluctuates in time. 
Thus the nature of the spectral line shape provide a useful information about the details of the 
dynamics of the solvent relative to the solute. An approximate treatment of the solvent dynamics 
allowed the theory to express the transition frequency fluctuation time correlation functions 
(related to the expressions for the absorption and emission line shapes) solely in terms of the 
two-body solute-solvent time-dependent conditional pair distribution function. 
Many other applications of pair dynamics have been discussed by a number of 
authors\cite{haan,balu1,posch,kivelson,rocca}.
    
The dynamics of a liquid below its freezing temperature, that is, 
in a supercooled state, is far more complex than what one would expect from a naive 
extrapolation of their high-temperature behavior. One of the most challenging problems in 
the dynamics of a supercooled liquid is to understand quantitatively the origin of the 
non-exponential relaxation exhibited by various dynamical response functions and the 
extraordinary viscous slow-down within a narrow temperature range as one approaches 
the glass transition temperature from above\cite{angell,arnab2}. Many experimental 
studies\cite{ediger,weitz} as well as computer 
simulations\cite{ka,glotzer,sastry,heuer} have been performed to shade light on 
the underlying microscopic mechanism involved in supercooled liquids. These studies have 
revealed the evidence of the presence of distinct relaxing domains (spatial heterogeneity) which 
is thought to be responsible for the non-exponential relaxations in deeply supercooled liquids. 
Molecular motions in strongly supercooled liquid involves highly collective movement of several
molecules\cite{glotzer,glotzer1,sarika1,sarika2,schober1}. Furthermore, the correlated jump motions 
become the dominant diffusive mode\cite{schober1,hansen}. The observed heterogeneity of the relaxations 
in a deeply supercooled liquid is found to be connected to the collective hopping of groups of 
particles\cite{schober2}. 

The occurrence of increasingly heterogeneous dynamics in supercooled liquids, however, has been 
investigated solely in terms of single particle dynamics. The study of the dynamics of pair of 
atoms which involve higher order (two-body) correlations thus can provide much broader insight 
into the anomalous dynamics of supercooled liquids. 
In this work, we have carried out molecular dynamics simulations in a strongly non-ideal
glass forming binary mixture (commonly known as Kob-Andersen model\cite{ka}) to study the relaxation 
mechanism in terms of the pair dynamics. The main purpose of the present study is to explore 
the dynamics in a more collective sense by following the relative motion of three different 
type ($AA$, $BB$, and $AB$) of nearest neighbor and next nearest neighbor pair of atoms.
These three pairs are found to behave differently. The 
simulation results show a clear signature of hooping motion in all the three pairs.
We have also performed simple mean-field (MF) model (as introduced by Haan\cite{haan} for 
one component liquid) calculations to obtain the time dependent conditional pair distribution 
functions. 

The organization of the rest of the paper is as follows. In Sec. II, we describe the details
of the simulation and the model system used in this study. The simulation results are
presented and discussed in Sec. III. In Sec. IV, we have presented a mean-field model
calculations for pair dynamics in binary mixture and the comparison is made with the
simulation results. Finally, we end with a few concluding remarks in Sec. V.

\section {System and Simulation Details}

We performed a series of equilibrium isothermal-isobaric ensemble (N P T) molecular 
dynamics (MD) simulation of a strongly non-ideal well-known glass forming binary 
mixture in three dimensions. The binary system studied 
here contains a total of $N = 1000$ particles consisting of two species of 
particles, $A$ and $B$ with $N_A = 800$ and $N_B = 200$ number of $A$ and $B$ 
particles, respectively. Thus, the mixture consists of $80 \%$ of A particles 
and $20 \%$ of B particles. The interaction between any two particles is modeled 
by means of shifted force Lennard-Jones (LJ) pair potential\cite{allen}, where the standard 
LJ is given by
\begin {equation}
u_{ij}^{LJ} = 4 \epsilon _{ij} \left[{\left(\sigma_{ij} \over r_{ij}\right)}^{12} - {\left(\sigma_{ij} \over r_{ij} \right)}^6 \right]
\end {equation}
\noindent where $i$ and $j$ denote two different particles ($A$ and $B$). 
The potential parameters are as follows: $\epsilon_{AA} = 1.0$, $\sigma_{AA} = 1.0$, 
$\epsilon_{BB} = 0.5$, $\sigma_{BB} = 0.88$, $\epsilon_{AB} = 1.5$, and $\sigma_{12} = 0.8$. 
The mass of the two species are same ($m_A = m_B = m$). Note that in this model system 
the $AB$ interaction ($\epsilon_{AB}$) is much stronger than both of their respective 
pure counterparts and $\sigma_{AB}$ is smaller than what is expected from the 
Lorentz-Berthelot mixing rules. In order to lower the computational burden the potential 
has been truncated with a cutoff radius of 2.5$\sigma_{AA}$. All the quantities in this study 
are given in reduced units, that is, length in units of $\sigma_{AA}$, temperature $T$ in 
units of $\epsilon_{AA}/k_B$, pressure $P$ in units of $\epsilon_{AA}/\sigma_{AA}^{3}$. 
The corresponding microscopic time scale is $\tau = \sqrt{m\sigma_{AA}^2/\epsilon_{AA}}$.

 All simulations in the NPT ensemble were performed using the 
Nose-Hoover-Andersen method\cite{nose} where the external reduced temperature 
is held fixed at $T^* = 1.0$. The external reduced pressure has been kept fixed 
at $P^* = 20.0$. The reduced average density ${\bar \rho}^*( = {\bar \rho}\sigma_{AA}^3)$ 
of the system corresponding to this thermodynamic state point is 1.32. Throughout the course 
of the simulations, the barostat and the system's degrees of freedom are coupled to an 
independent Nose-Hoover chain\cite{martyna1} (NHC) of thermostats, each of length 5. 
The extended system equations of motion are integrated using the reversible 
integrator method\cite{tuckerman} with a time step of $0.002$. The higher order 
multiple time step method has been employed in the NHC evolution operator 
which lead to stable energy conservation for non-Hamiltonian dynamical systems\cite{martyna2}. 
The extended system time scale parameter used in the calculations was taken to be $1.15$ 
for both the barostat and thermostats.   
   
 The systems were equilibrated for $2 \times 10^6$ time steps and simulations were 
carried out for another $10^7$ production steps following equilibration, during 
which the quantities of interest are calculated.

\section{Simulation Results and Discussion} 

The three partial radial distribution functions, $g_{AA}(r)$, $g_{AB}(r)$ and $g_{BB}(r)$
obtained from simulations are plotted in figure 1. Due to the strong
mutual interaction, the AB correlation is obviously the strongest among the three pairs. The splitting 
of the second peak of both $g_{AA}(r)$ and $g_{AB}(r)$ is 
a characteristic signature of dense random packing. The structure of $g_{BB}(r)$ is 
interestingly different. It has an insignificant first peak which originates from the weak 
interaction between the B type particles. The second peak of $g_{BB}(r)$ is higher than 
that of the first peak signifying that the predominant BB correlation takes place at the 
second coordination shell. The occurrence of the splitted second peak is clearly observed here 
also.    

In the present study, the central quantity of interest is the {\it time-dependent pair 
distribution function} (TDPDF) (first introduced by Oppenheim and Bloom\cite{bloom} in the theory of
nuclear spin relaxation in fluids), $g_2({\bf r}_o,{\bf r};t)$ which is the conditional
probability that two particles are separated by ${\bf r}$ at time $t$ if that pair were 
separated by ${\bf r}_o$ at time $t = 0$. Thus, the TDPDF measures the relative motion of 
a pair of atoms. For an isotropic fluid, the TDPDF depends only on the magnitudes 
of ${\bf r}$, ${\bf r}_o$ and $\theta$, where $\theta$ is the angle between ${\bf r}$ and 
${\bf r}_o$. In computer simulation, one can readily evaluate separately the radial and 
orientational features of the relative motion. In the following two subsections, we present, 
respectively, the results obtained for the time evolution of the radial part $g_{2,rad}(r_o,r;t)$ 
and the angular part $g_{2,ang}(r_o,\theta;t)$ of the TDPDF for the three different 
pairs (AA, BB and AB).

\subsection{Radial part of the TDPDF, $g_{2,rad}(r_o,r;t)$}

In figure 2 we plot the $g_{2,rad}(r_o,r;t)$ for the AA pair with the 
initial separation $r_o$ corresponds to the first maximum of the partial radial
distribution function $g_{AA}(r)$ (that is, pairs which are nearest neighbor) at 
four different times. While at short time (figure 2(a)) the distribution function has a 
single peak structure as expected, with increase in time it reaches slowly to its 
asymptotic limit where additional peaks develop at larger relative separations 
(see figures 2(b)-2(d)). The microscopic details of the underlying diffusive process
(by which it approach to the asymptotic structure) can be obtained by following the trajectory of
the relative motions. Figure 3 displays the projections onto an x-y plane of the 
trajectory of a typical AA pair for the nearest neighbor A atoms over a time 
interval $\Delta t = 500\tau$. The motion 
of the AA pair is shown to be relatively localized for many time steps and then move
significant distances only during quick, rare cage rearrangements. This is a clear evidence that
the jump motions are the dominant diffusive mode by which the separation between pairs of atoms 
evolve in time. 

The behavior of the distribution function $g_{2,rad}(r_o,r;t)$ for the AB pairs (where the 
interaction being the strongest) is plotted in figure 4 at four different times. The  
distribution function shows the same qualitative behavior as we observed in the case 
of AA correlation (figure 2). When compared to the 
AA correlation function within the same time scale, the decay of the correlation function is found
to be faster despite the much strong AB interaction. This must be attributed to the difference in 
size of the two types of particles. As the B particles are smaller in size than the A particles, they
are more mobile. In addition, the AB interaction is such that AB repulsion is felt at relatively 
small distances ($\sigma_{AB}$ = 0.8 instead of 0.94 according to the Lorentz-Berthelot rules). 
Consequently, the B particles are more prone to make jumps than the A particles (as observed earlier
by Kob and Andersen\cite{ka}).
     
The nature of the relative motion of a typical AB pair is illustrated in figure 5(a), which display 
the trajectory of a typical AB pair (in the x-y plane) that were 
initially at the nearest neighbor (first peak of $g_{AB}(r)$). The elapsed time 
is $\Delta t = 500\tau$. The dynamics of the relative motion is again
dominated by hooping, the AB pair remain trapped at their initial separation over 
hundreds of time steps, before jumping to neighboring sites where they again become 
localized. Further, the jump motion is more frequent
for the AB pair than that for the AA pair. The individual trajectory of the A and B particles 
of the same AB pair within the same time window is shown in figures 5(b) and 5(c), respectively. 
While both the A and B particles hop, B particles move faster and 
the effect of caging is weaker (than the A particles) due to its smaller size. In this time window, 
the net displacement of the AB pair in the x-y plane is found to be quite large as shown 
in figure 6 and mainly determined by the displacement of the B particle.

In figure 7 we show $g_{2,rad}(r_o,r;t)$ for the BB pair initially separated at the 
first peak of $g_{BB}(r)$ at four different times. Due to weak interaction among B particles, one 
expects that the B atoms in the BB pair 
will fast lose the memory of their initial separation. This is indeed the case for the
BB pair shown in figure 7. Once again the jump dynamics is clearly seen in the 
trajectory of a typical BB pair projected in the x-y plane (figure 8).  

We now consider the case where the separation of the initial pairs corresponds to the 
second peak of their respective partial radial distribution functions of figure 1 (that is, 
pairs which are next nearest neighbor). The distribution function for the AA pair is plotted 
in figure 9. It shows a qualitative different behavior because the peak at the nearest neighbor 
separation develops in a relatively short time. Here also 
the motion of the pairs are found to be mostly discontinuous in nature, thus motion from second to
first nearest neighbor occurs mostly by hopping. 
In figure 10 we plot the similar distribution function of 
the AB pair. Since the AB interaction is the strongest, the height of the 
first peak grows faster than that for the AA pair (compare figures 9(b) and 10(b)). 
Next, in figure 11 we plot the distribution function for the BB pair. Contrary to the AA and AB pairs,
BB pairs tend to retain their initial separation for a relatively long time compared to the nearest
neighbor pair. This can be understood 
from the predominant BB correlations at the second coordination shell.

\subsection{Angular part of the TDPDF, $g_{2,ang}(r_o,\theta;t)$}

In this subsection we present the {\it angular distribution function} $g_{2,ang}(r_o,\theta;t)$
for the three different pairs (AA, BB and AB). The initial separation $r_o$ for the three
pairs corresponds to the first peak of the respective partial radial distribution
functions (figure 1).
  
In figure 12(a) we show the angular distribution $g_{2,ang}(r_o,\theta;t)$ for the AA pair.
We calculate the angular distribution with respect to the initial separation vector 
${\bf r_o}$ and irrespective of the value of the separation at time t. The distribution
which is a $\delta$-function at $t = 0$, spreads more and more with time and eventually 
it reaches to a uniform distribution with zero slope. When we compare to the distribution 
corresponding to the AB pair as shown in figure 12(b), we find that the approach to the uniform
value is faster in the case of AB pair. This can be understood again in terms of the 
mobility of the B particles which is more compared to the A particles. In figure 12(c) we
show how the distribution for the BB pair changes with time. The relaxation
is seen to be relatively slower at short times compared to the AB pair. This can be understood in 
terms of the effective potential, discussed later.

\subsection{Relative diffusion: Mean square relative displacement (MSRD)}

In this subsection, we investigate the time dependence of the mean square 
relative displacement $\langle\mid{\bf r}_{ij}(t) - {\bf r}_{ij}(0)\mid^2\rangle_{r_o}$, the simplest 
physical quantity associated with the pair motion, where the
index $i$ and $j$ denote A and/or B particles and the subscript ${r_o}$ indicate the ensemble 
averaging is restricted to the pairs whose initial separation corresponds to $r_o$\cite{balu2}.  
First, we consider the case where the initial separations for the three pairs corresponds to the
first peak of the respective partial radial distribution functions (see figure 1). 
In other words, we consider first those pairs that were initially nearest neighbor pairs.

Figure 13 shows the result for the time dependence of the relative mean
square displacement (MSRD) of the three pairs. At long times
the MSRD varies linearly with time. However, the evolution
of MSRD with time differs for different pairs.
As expected, the smaller size of the B particles and the weak BB interaction
leads to a faster approach of the diffusive limit of BB pair separation. The time scale needed 
to reach the diffusive 
limit is shorter for the AB pair than that for the AA pair.

From the slope of the curves in the linear region one can obtain the values 
of the relative diffusion constants $D_R$ of the different pairs. The values thus 
obtained are the following: $D_R^{AA} \simeq 0.0032$, $D_R^{AB} \simeq 0.0048$ and
$D_R^{BB} \simeq 0.0064$. One should note that even though the difference in size of the A and
B particles is small, $D_R^{BB}$ is almost twice of $D_R^{AA}$.  
At sufficiently long time, one would certainly expect
the diffusion constant for the relative motion of a pair should be the sum of the
individual diffusion constants of the two atoms obtained from the slope of the 
corresponding mean square displacements at long time. Indeed, we find there is a 
good agreement.  
  
An investigation of the behavior of MSRD is also performed for atomic pairs which were
initially next nearest neighbor. When compared to the nearest neighbor
pairs (figure 13) we find that the slope of the corresponding straight lines 
are almost identical, although in the case of AA and AB pairs the diffusive limits are reached 
at shorter times. This has been shown in figure 14. One should remember
that the AA and AB correlations are highest at the first coordination shell whereas the 
highest BB correlations occur at the second coordination shell (see figure 1). Thus, 
at short time the increase in slope for the AA and AB pairs can be 
explained in terms of the decrease in correlations at the second coordination shell.

\subsection{The non-Gaussian parameter for the relative motion}

In a highly supercooled liquid, the single particle 
displacement distribution function $G_s(r,t)$ (known as the self-part of the 
van Hove correlation function) has an extended tail and
is, in general, non-Gaussian. The deviation from the Gaussian behavior
is often thought to reflect the presence of the transient inhomogeneities
and can be characterized by the 
non-Gaussian parameter $\alpha_2(t)$\cite{glotzer}
\begin{equation}
\alpha_2(t) = {3 \langle\Delta r^4(t)\rangle\over 5 \langle\Delta r^2(t)\rangle^2} - 1,
\end{equation}
\noindent
where $\langle\Delta r^2(t)\rangle$ and $\langle\Delta r^4(t)\rangle$ are the second and fourth 
moments of $G_s(r,t)$, respectively.
At intermediate time scale, $\alpha_2(t)$ increases with time and the maximum of 
$\alpha_2(t)$ occurs around the end of the $\beta$ relaxation region. Only on the time 
scale of diffusion or the $\alpha$ relaxation, $\alpha_2(t)$ starts to decrease and 
finally at very long time limit, it reaches to zero. 
$\alpha_2(t)$ calculated for the A and B particles are shown 
in figure 15. The maximum in $\alpha_2(t)$ is seen to shift slightly towards left and 
also the height of the maximum is higher for the B particles. This is a clear evidence
that the B particles probe much
more heterogeneous environment than does the A particles. This difference can be explained 
in terms of the smaller concentration of B particles, different sizes of the A and B particles 
and also from the fact that the interaction
between the B particles is much weaker than that between the A particles\cite{ka,glotzer}.

Motivated by these findings for the single particle displacement distribution 
function, we introduce a new non-Gaussianity parameter for the pair dynamics, denoted by
$\alpha_2^P(t)$. $\alpha_2^P(t)$ can be a measure of the deviation from the Gaussian
behavior of the pair distribution function $g_2({\bf r}_o,{\bf r};t)$. It can be 
defined as,
\begin{equation}
\alpha_2^{P_{ij}}(t) = {3 \langle\mid{\bf r}_{ij}(t) - {\bf r}_{ij}(0)\mid^4\rangle_{r_o}\over 5 \langle\mid{\bf r}_{ij}(t) - {\bf r}_{ij}(0)\mid^2\rangle_{r_o}^2} - 1, \,\,\,\, (i,j = A\,and/or\,B)
\end{equation}
\noindent
where $\langle\mid{\bf r}_{ij}(t) - {\bf r}_{ij}(0)\mid^2\rangle_{r_o}$ and 
$\langle\mid{\bf r}_{ij}(t) - {\bf r}_{ij}(0)\mid^4\rangle_{r_o}$
are the mean square relative displacement and mean quartic relative displacement of the 
$ij$ pair. One should note that $\alpha_2^P(t)$ is identical to zero for a Gaussian 
pair distribution function. 

In figure 16 we show the behavior of the $\alpha_2^{P_{ij}}$ as a function of time for
the three different pairs. We again consider only those pairs that were initially nearest neighbor.
The behavior observed for the three pairs is markedly different. 
The dynamics explored by the BB pair 
is seen to be less heterogeneous than the AA and AB pairs. Because of the smaller size of the B particles, the B particles reach the average distribution faster, although it explores larger heterogeneity. 
The AA pair reaches 
the diffusive limit at longer time scale than that for the AB pair, the AB pair
explore more heterogeneous dynamics as is clearly evident from the difference in the
maximum value of $\alpha_2^P(t)$.

\section{Theoretical Analysis}

For the motion of an atomic pair in a pure fluid, Haan\cite{haan} introduced a simple 
mean-field level equation of motion for the time-dependent pair distribution function $g_2$.
This equation was shown to give a quantitatively correct description both at short and long times\cite{skinner}. 
This treatment is mean-field in the sense that the two atoms were
assumed to diffuse in an effective-force field of the surrounding particles given by
the gradient of the potential of mean force. The equation for $g_2$ was represented by a 
Smoluchowski equation and the correct short time description of $g_2$ was obtained only by 
introducing a {\it nonlinear time that is related to the mean-squared distance} (MSD) moved 
by a {\it single atom}. In other
words, an {\it ad hoc} introduction of a time-dependent
diffusion constant $D(t)$ in the equation of motion gives the correct description at short times. 

 In the view of its success for one-component liquid, we have performed a similar mean-field 
model calculations for the binary mixture considered here. The generalization 
to binary mixture gives the following Smoluchowski equation for the different pairs
\begin{equation}
{\frac{\partial g_2^{ij}({\bf r_o},{\bf r};t)}{\partial \tau_{ij}}}\;=\;\nabla \cdot \biggr[\nabla g_2^{ij}({\bf r_o},{\bf r};t) + \beta g_2^{ij}({\bf r_o},{\bf r};t)\nabla W_{ij}(r)\biggl], 
\end{equation}
\noindent
where the index $i$ and $j$ denote A and/or B particles. $\beta$ is the inverse of the Boltzmann's 
constant ($k_B$) times the absolute temperature (T).
$W_{ij}(r)$ is the potential of mean force ('effective potential') between the $i$ and $j$ particles 
given by
\begin{equation}
W_{ij}(r)\;=\;-k_BT \ln g_{ij}(r)
\end{equation}
\noindent 
where $g_{ij}(r)$ is the partial radial distribution function. In Eq. 4, the 
"time" $\tau_{ij}$ is defined by
\begin{eqnarray}
\tau_{ij}&=&{\frac{1}{6}}\langle\mid{\bf r}_{ij}(t) - {\bf r}_{ij}(0)\mid^2\rangle_{r_o} \nonumber\\
&&\approx{\frac{1}{6}}\biggr[\langle\mid{\bf r}_i(t) - {\bf r}_i(0)\mid^2\rangle + \langle\mid{\bf r}_j(t) - {\bf r}_j(0)\mid^2\rangle
\biggl]
\end{eqnarray}
\noindent
where $\langle\mid{\bf r}_{ij}(t) - {\bf r}_{ij}(0)\mid^2\rangle_{r_o}$ is the mean square relative 
displacement (MSRD) 
of the $'ij'$ pair. Note that an approximation is made in the above equation by neglecting the 
cross-correlation between the two particles ('i' and 'j') and the MSRD is replaced by the sum 
of individual particle's mean square displacement (MSD). 

Now the integration of the $g_2^{ij}({\bf r}_o,{\bf r};t)$ over the solid angles $\hat{\Omega_o}$ and 
$\hat{\Omega}$ corresponding to the initial and final positions, respectively, gives the radial part 
of the full distribution function (the zeroth-angular moment of $g_2^{ij}({\bf r_o},{\bf r};t)$)
\begin{equation}
g_{2,rad}^{ij}(r_o,r;t)\;=\;{\frac{1}{4\pi}}\,\int\,d\hat{\Omega_o}\,d\hat{\Omega}\,g_2^{ij}({\bf r_o},{\bf r};t)
\end{equation}
\noindent
Note that the normalization of this function is
\begin{equation}
\int_0^{\infty}\,dr\,r^2\,g_{2,rad}^{ij}(r_o,r;t)\;=\;1
\end{equation} 
The equation of motion for $g^{ij}_{2,rad}(r_o,r;t)$ (derived from Eq. 4) is solved numerically 
(by Crank-Nicholson method) for the different pairs and the results obtained from this model 
calculations are compared with the simulation results. The partial radial distribution 
functions $g_{ij}(r)$ and the mean-square displacements of the A and B particles (required as 
input) are obtained from the present simulation. 

Figures 17 and 18 compare model calculations with the simulated distribution functions for the AA 
and AB nearest neighbor pairs. The time evolution of the distributions 
are described well by the simple mean-field model. The underlying effective-potential energy 
surfaces are plotted in figure 19. Thus relative diffusion in these cases can be 
considered as overdamped motion in an effective potential, which takes place mainly via 
hopping (as shown in figures 3 and 5), which governs the time evolution of the distributions for 
the AA and AB pair. 

Unfortunately, the good agreement observed above between simulation and theory for the AA and the 
AB pairs, does not extend to the BB pair. This is shown in figure 20. As the number of B particles 
present in the system is much less (20 \%) and the interparticle interaction is weak, the effective 
potential for a B atom interacting with a nearest neighbor atom is unfavorable (see the figure 19). 
Consequently, the nearest neighbor BB pairs execute highly anharmonic motions.
Thus, the fluctuations about the mean-force field experienced by the BB pair are large and important.
These fluctuations are neglected here, as in other mean-field level description. 

The extension of the calculations to the case of next nearest neighbor pairs also been carried
out and compared with the simulated distributions. It should be noted that compared to the 
nearest neighbor pairs, the AA and AB pairs are now executes motions in a relatively weak, shallow 
potential, whereas the motions of the BB pairs takes place in a relatively strong, bound potential 
well (see the figure 19). Thus, for the BB pairs, one expects a better agreement with the simulated 
distributions compared to the earlier case (nearest neighbor BB pairs). Indeed, the agreement is 
better for the BB pairs, as shown in figure 21 (compare with figure 20). We have found that 
the MF model provides a good description of the dynamics of the AA and AB pairs, although the 
agreement is not as satisfactory as for the nearest neighbor pairs.

Thus, it is evident that the MF description for the time dependent pair distribution functions is 
reasonably good for the AA and AB pairs. Simulation results have shown that the relative diffusion 
of an AB pair is higher than that for an AA pair. We noted that this due to more frequent hopping of B particles than the A particles. Our main objective now is to see whether the frequent
jump motions of the B particles, as predicted by the simulations, can be explained in terms of the 
MF model described above. 

We have performed an approximate calculation to get 
an estimate of the transition rate between the first two adjacent minima in the effective potential 
energy surface of the AA and AB pairs (see the figure 21). In other words, the rate of crossing
from the deep minima located at the nearest neighbor pairs, to the adjacent minima (corresponds
to the next nearest neighbors). As the motion of a pair in the effective potential was treated
by a Smoluchowski equation, we use the corresponding rate expression in the overdamped limit to
calculate the escape rate. Thus, we have an expression for the escape rate given by\cite{zwanzig}
\begin{equation}
k_{S}\;\cong\;{\frac{\omega_{min}\omega_{max}}{2\pi\zeta}}\exp\Biggl(-{\frac{\Delta W}{k_BT}}\Biggr)
\end{equation}
\noindent
where $\Delta W = W(r_{max}) - W(r_{min})$ is the Arrhenius activation energy and $\omega_{min}$, 
$\omega_{max}$ are the frequency at the minima ($r_{min}$) and maxima ($r_{max}$) in the 
effective potential $W(r)$, respectively. The diffusion coefficient $D$ is related to the friction 
$\zeta$ by $D = k_BT/\zeta$.

Thus, to calculate the transition rate we need to know the values of the frequencies $\omega_{min}$, 
$\omega_{max}$, and the barrier height $\Delta W$, which are different for the AA and AB pairs.
For the AA pairs, these parameters are found to be $\omega_{min}^* \simeq 16.5$, 
$\omega_{max}^* \simeq 6.5$ and $\Delta W_{AA} \simeq 2.25 k_BT$, whereas for the AB pairs they are 
$\omega_{min}^* \simeq 17.8$, $\omega_{max}^* \simeq 7.4$ and $\Delta W_{AB} \simeq 2.45 k_BT$.
The relative diffusion of the two pairs are $D_R^{AA} \simeq 0.0032$ and $D_R^{AB} \simeq 0.0048$.
Using all these parameters, the escape rate calculated for the AA and AB pairs are (in reduced units) 
$k_S^{AA} \simeq 5.9 \times 10^{-3}$ and $k_S^{AB} \simeq 8.8 \times 10^{-3}$, respectively (in terms of time $\tau$, which is equal to 2.2 ps for argon units).

Even though the barrier height $\Delta W_{AB} > \Delta W_{AA}$, the transition rate for the
AB pair is larger than that for the AA pair. Thus, the jump motions are much more frequent for the 
AB pair, due to the large diffusion of the B particles in the potential
energy surface (which mainly occurs via hopping mode).

\section{Conclusions}

 Let us first summarize the main results of this study. We have presented the molecular dynamics
simulation results for the time dependent pair distribution functions in a strongly non-ideal
glass forming binary Lennard-Jones mixture. In addition, a mean-field description of the pair
dynamics is considered and the comparison is made with the simulated distributions. The main goal
of this investigation was to explore the dynamics of the supercooled liquids in a more collective
way, by following the {\it relative} motion of the atoms rather than {\it absolute} motion 
of the atoms. We find that the three pairs (AA, BB and AB) behave differently. The analysis of
the trajectory shows a clear evidence of the jump motions for all the three pairs. 

The relative diffusion constant of the BB pair ($D_R^{BB}$) is almost twice the value for the
AA pair ($D_R^{AA}$). This clearly suggests the importance of the jump dynamics for the B particles
and indeed, we find that the motion of the B particles is mostly discontinuous in nature, while
the A particles show occasional hopping. The dynamic inhomogeneity present in a supercooled liquid is
generally characterized by the well-known non-Gaussian parameter $\alpha_2(t)$, which describe
the deviations from the Gaussian behavior in the motion of a single atom. In this paper, we 
have generalized this concept and introduce a non-Gaussian parameter for the pair dynamics ($\alpha_2^P(t)$), to measure the deviations from the Gaussian behavior in the relative motion of the atoms.
At intermediate times, all the three pair distribution functions for the three pairs show significant 
deviations from the Gaussian behavior, with different degree. 

It is found that for the nearest neighbor AA and AB pairs, which are confined to a strong effective 
potential and merely makes anharmonic motions in it, the dynamics can be treated at the mean-field 
level. However, as the motion of a nearest neighbor BB pair is highly anharmonic, one must
include the effects of the fluctuations about the mean-force field, in order to get a correct
description of the dynamics. 

 While the mean-field treatment provides reasonably accurate description of pair dynamics (at least
for AA and AB pairs), it must be supplemented with the time dependent diffusion coefficient (D(t)). 
This is a limitation of the mean-field approach because at present we do not have any theoretical means
to calculate D(t) from first principles. The mode coupling theory (MCT) does not work because it 
neglects hopping which is the dominant mode of mass transport in deeply supercooled liquids, even when 
the system is quite far from the glass transition. As we discussed recently, the hopping can be coupled
to anisotropy in the local stress tensor\cite{sarika1}. The calculation of the latter is also 
non-trivial. Work in this direction is under progress.

\vspace{1cm}
{\bf Acknowledgments}

This work was supported in part by the Council of Scientific and Industrial
Research (CSIR), India and the Department of Science and Technology (DST), India.
One of the authors (R.K.M) thanks the University Grants Commission (UGC) for
providing the Research Scholarship.

\newpage
 
{\large \bf Figure Captions}

{\bf Figure 1.} The radial distribution function $g(r)$ for the AA, AB, and BB 
correlation is plotted against distance. The solid line is $g_{AA}(r)$, the dashed line
is $g_{AB}(r)$, and the dot-dashed line is $g_{BB}(r)$. For details, see the text.

{\bf Figure 2.} The radial part of the time-dependent pair distribution function $g_{2,rad}(r_o,r;t)$
for the AA pair as a function of separation $r$ at four different times: (a) $t = 20\tau$, (b) $t = 50\tau$, (c) $t = 100\tau$, and (d) $t = 300\tau$. The initial separation $r_o$ 
corresponds to the first maximum of $g_{AA}(r)$. Note that the time unit $\tau = 2.2 ps$ if argon units
are assumed.   

{\bf Figure 3.} Projections into x-y plane of the trajectory of a typical nearest neighbor AA pair 
over a time interval $t = 500\tau$. Note that the time unit $\tau = 2.2 ps$ for argon units.

{\bf Figure 4.} The radial part of the time-dependent pair distribution function $g_{2,rad}(r_o,r;t)$ for the AB pair as a function of separation $r$ at four different times: (a) $t = 20\tau$, (b) $t = 50\tau$, (c) $t = 100\tau$, and (d) $t = 300\tau$. The initial separation $r_o$ corresponds to the first maximum of $g_{AB}(r)$. The time unit $\tau = 2.2 ps$ for the argon units. For further details, see the text. 

{\bf Figure 5.} (a) Projections into x-y plane of the trajectory of a typical nearest neighbor AB pair over a time interval $t = 500\tau$. (b) Trajectory of the A particle of the same AB pair as in (a), within 
the same time window. (c) Trajectory of the B particle of the same AB pair.       
   
{\bf Figure 6.} The net displacement of an AB pair into x-y plane ($\Delta L^{xy}$) as shown in 
figure 5(a), in the same time interval. Note that the displacement is quite large. 

{\bf Figure 7.} The radial part of the time-dependent pair distribution function $g_{2,rad}(r_o,r;t)$ for the BB pair as a function of separation $r$ at four different times: (a) $t = 20\tau$, (b) $t = 50\tau$, (c) $t = 100\tau$, and (d) $t = 300\tau$. The initial separation $r_o$ corresponds to the first peak
of $g_{BB}(r)$. 

{\bf Figure 8.} Projections into x-y plane of the trajectory of a typical nearest neighbor BB pair over
a time interval $t = 500\tau$.    

{\bf Figure 9.} The radial part of the pair distribution function $g_{2,rad}(r_o,r;t)$ for the AA 
pair at four different times: (a) $t = 4\tau$, (b) $t = 20\tau$, (c) $t = 100\tau$, and (d) $t = 300\tau$.
Here the initial separation $r_o$ is chosen at the second peak of $g_{AA}(r)$. 

{\bf Figure 10.} The radial part of the pair distribution function $g_{2,rad}(r_o,r;t)$ for the AB pair at four different times: (a) $t = 4\tau$, (b) $t = 20\tau$, (c) $t = 100\tau$, and (d) $t = 300\tau$.
Here the initial separation $r_o$ corresponds to the second peak of $g_{AB}(r)$. 

{\bf Figure 11.} The radial part of the pair distribution function $g_{2,rad}(r_o,r;t)$ for the BB pair at four different times: (a) $t = 4\tau$, (b) $t = 20\tau$, (c) $t = 100\tau$, and (d) $t = 300\tau$.
Here the initial separation $r_o$ is chosen at the second peak of $g_{BB}(r)$. 

{\bf Figure 12.} (a) The angular part of the time-dependent pair distribution function $g_{2,ang}(r_o,\theta;t)$ for the AA pair at four different times. (b) $g_{2,ang}(r_o,\theta;t)$ for the AB pair. (c) $g_{2,ang}(r_o,\theta;t)$ for the BB pair. In all the three cases, we consider only those pairs
which were initially separated at the nearest neighbor distance. For further details, see the text.

{\bf Figure 13.} Time dependence of the mean square relative displacement (MSRD) for the AA, AB and BB pair (in units of $\sigma_{AA}^2$). The initial separation $r_o$ of the three pairs corresponds to the first 
peak of the respective partial radial distribution functions. The solid line represents the result for 
the AA pair, the dashed line AB pair, and the dotted line for the BB pair. 
For the detailed discussion, see the text.

{\bf Figure 14.} (a) Comparison of the mean square relative displacement (MSRD) for the AA pair with different initial separations. The solid line represents the nearest neighbor AA pair and the dashed line represents the next nearest neighbor AA pair. (b) Same as in (a), but for the AB pair. (c) For the BB pair. For details, see the text.  

{\bf Figure 15.} The behavior of the non-Gaussian parameter $\alpha_2(t)$ as a function of time for 
the A and B particles. The solid line is for the A particles and the dashed line for the B particles.

{\bf Figure 16.} The behavior of the non-Gaussian parameter $\alpha_2^P(t)$ as a function of time for the AA, AB, and BB pairs, initially separated at the nearest neighbor distance. The solid line represents the result for the AA pair, the dashed line for the AB pair, and the dot-dashed line for the BB pair.

{\bf Figure 17.} The simulated distribution $g_{2,rad}(r_o,r;t)$
for the AA pair is compared with the mean-field model calculations at three different times: (a) $t = 10\tau$, (b) $t = 50\tau$, and (c) $t = 100\tau$. The initial separation $r_o$ 
corresponds to the first maximum of $g_{AA}(r)$. The histogram represents the simulation results and the dashed line represents the results of the model calculations. Note that the time unit $\tau = 2.2 ps$ if 
argon units are assumed.  

{\bf Figure 18.} The simulated distribution $g_{2,rad}(r_o,r;t)$
for the AB pair is compared with the mean-field model calculations at three different times: (a) $t = 10\tau$, (b) $t = 50\tau$, and (c) $t = 100\tau$. The initial separation $r_o$ 
corresponds to the first maximum of $g_{AB}(r)$. The histogram represents the simulation results and the dashed line represents the results of the model calculations. 

{\bf Figure 19.} The potential of mean force $W(r)$ for the AA, AB and BB pairs in the Kob-Andersen model at the reduced pressure $P^* = 20$ and the reduced temperature $T^* = 1.0$. The solid line represents
for the AA pair, the dashed line for the AB pair, and the dot-dashed line for the BB pair.

{\bf Figure 20.} The simulated distribution $g_{2,rad}(r_o,r;t)$
for the BB pair is compared with the mean-field model calculations at three different times: (a) $t = 10\tau$, (b) $t = 50\tau$, and (c) $t = 100\tau$. The initial separation $r_o$ 
corresponds to the first peak of $g_{BB}(r)$. The histogram represents the simulation results and the dashed line represents the results of the model calculations. 

{\bf Figure 21.} The simulated distribution $g_{2,rad}(r_o,r;t)$
for the BB pair is compared with the mean-field model calculations at two different times: (a) $t = 10\tau$and (b) $t = 100\tau$. Here the initial separation $r_o$ 
corresponds to the second peak of $g_{BB}(r)$. The histogram again represents the simulation results and the dashed line represents the results of the model calculations. For further details, see the text.
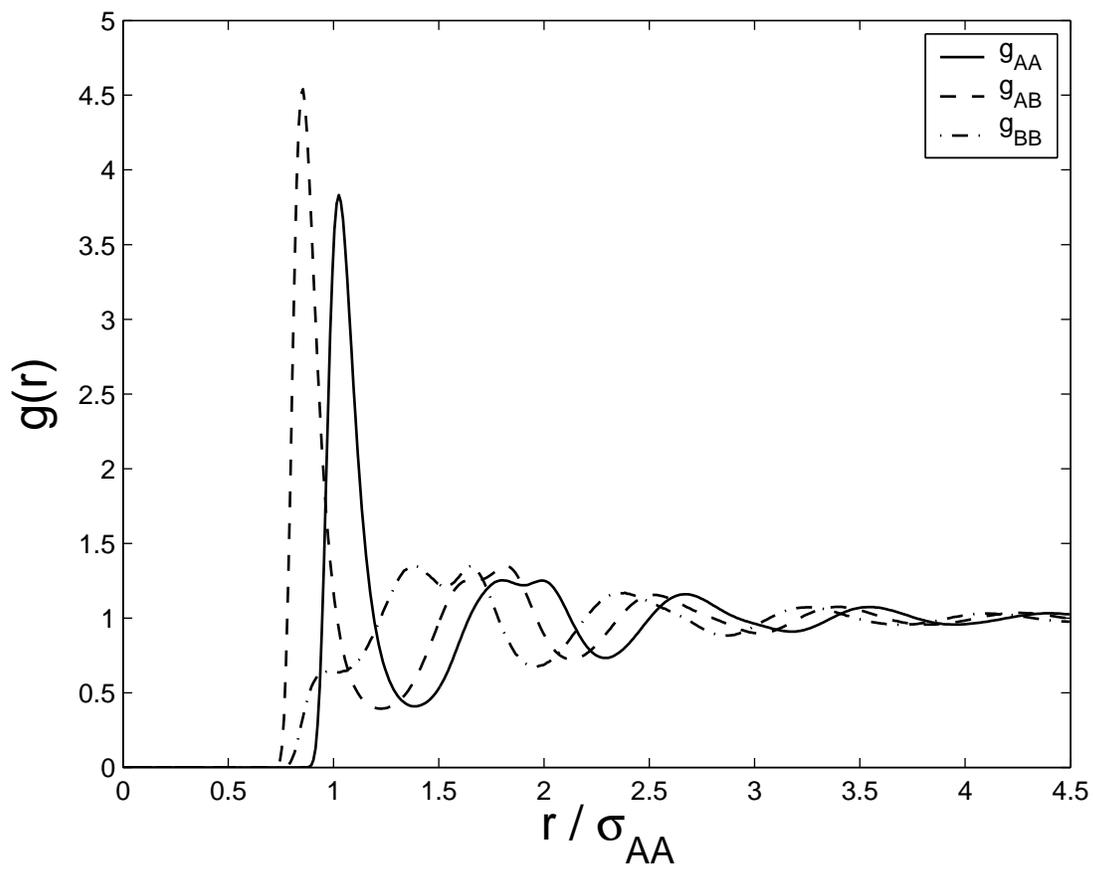
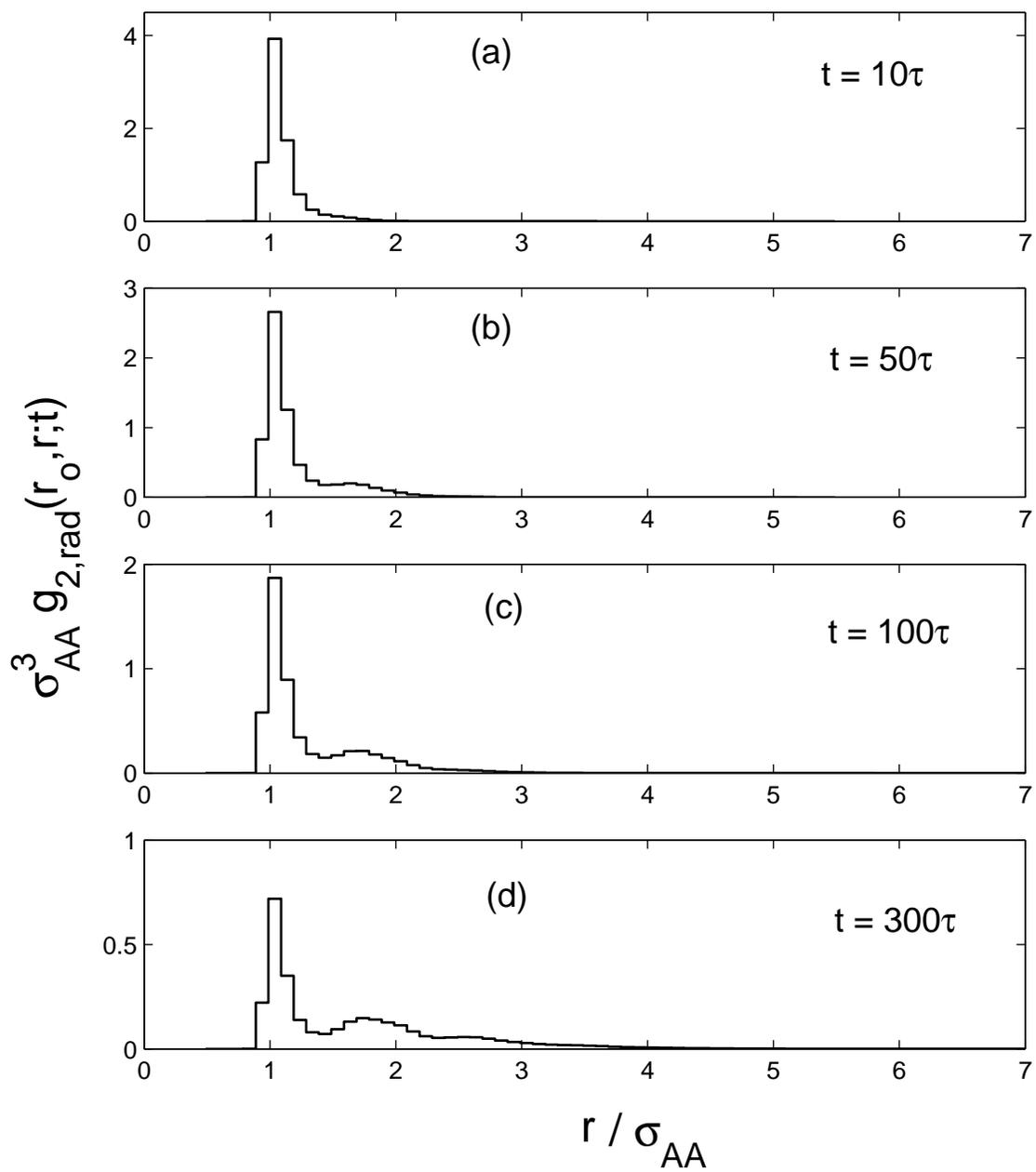
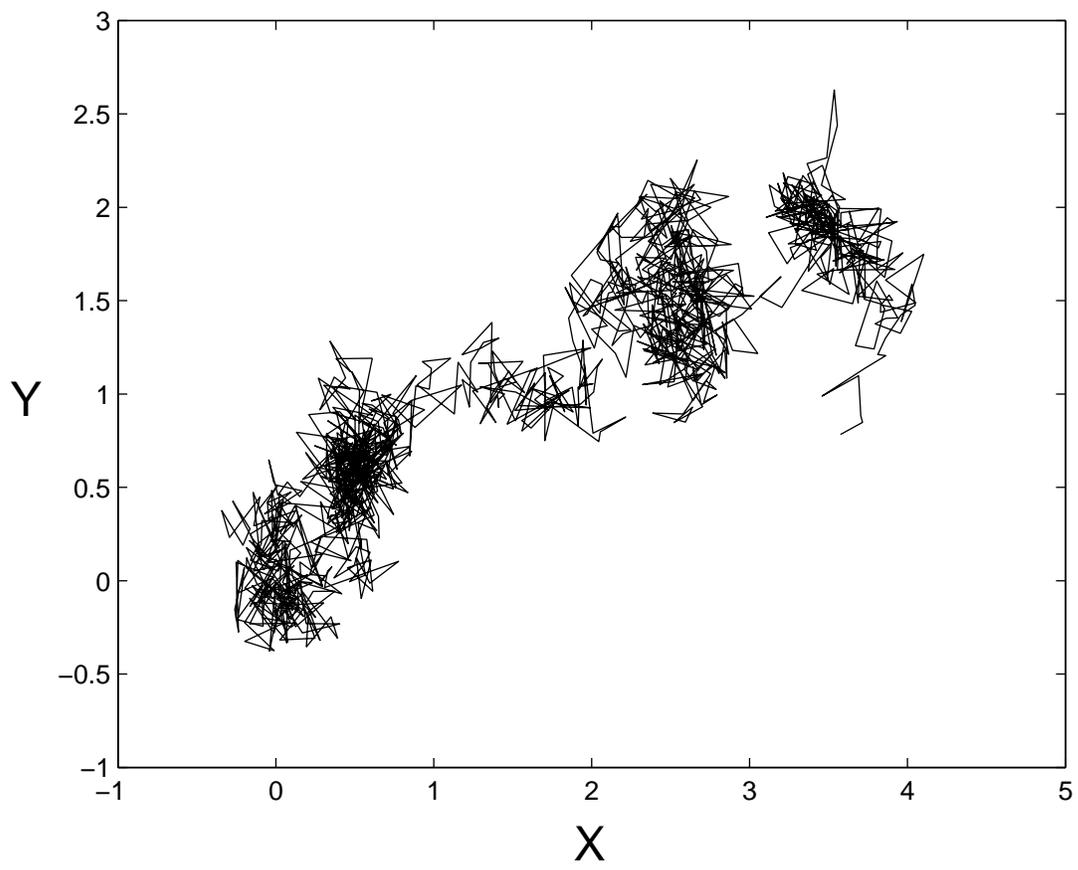
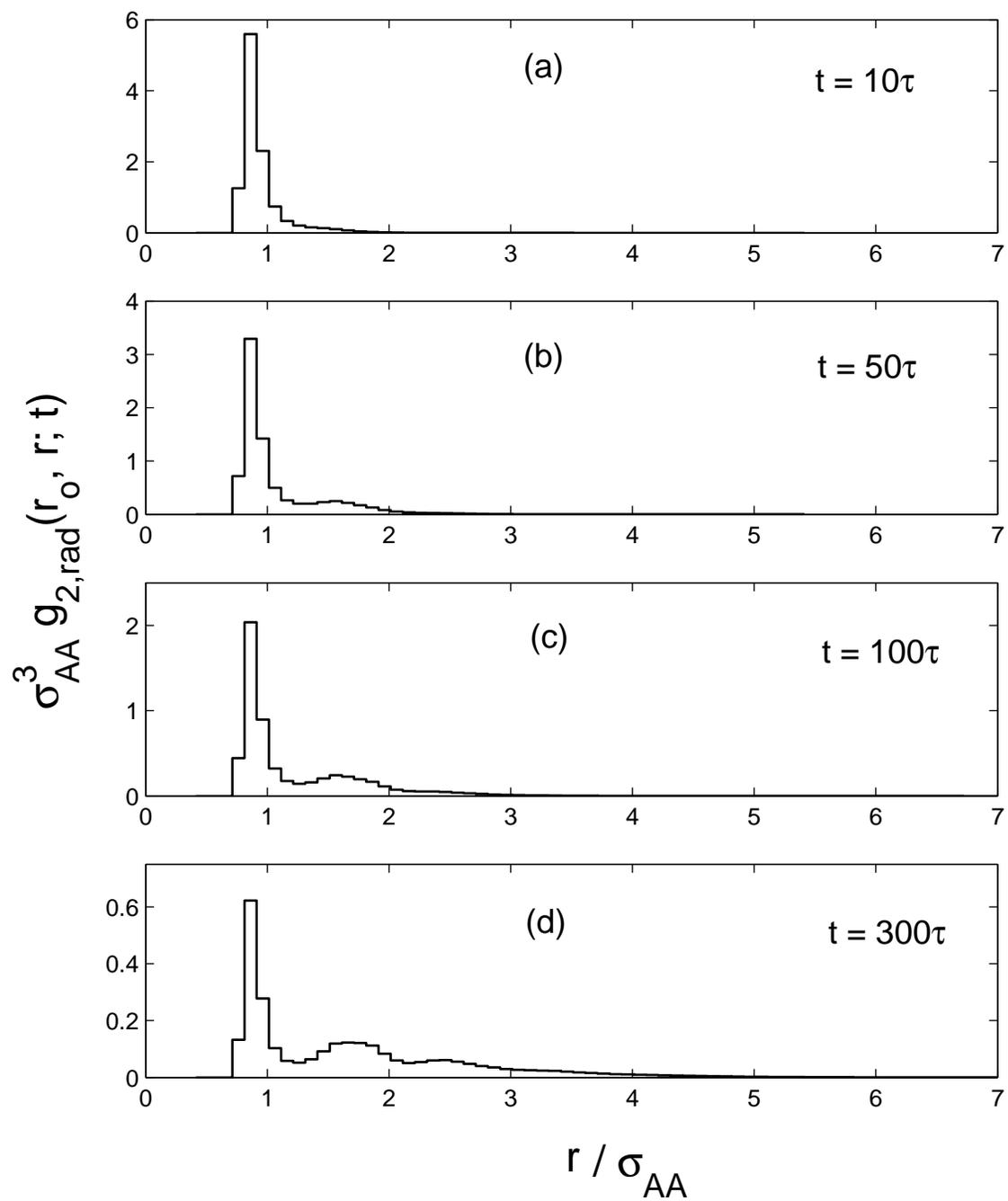
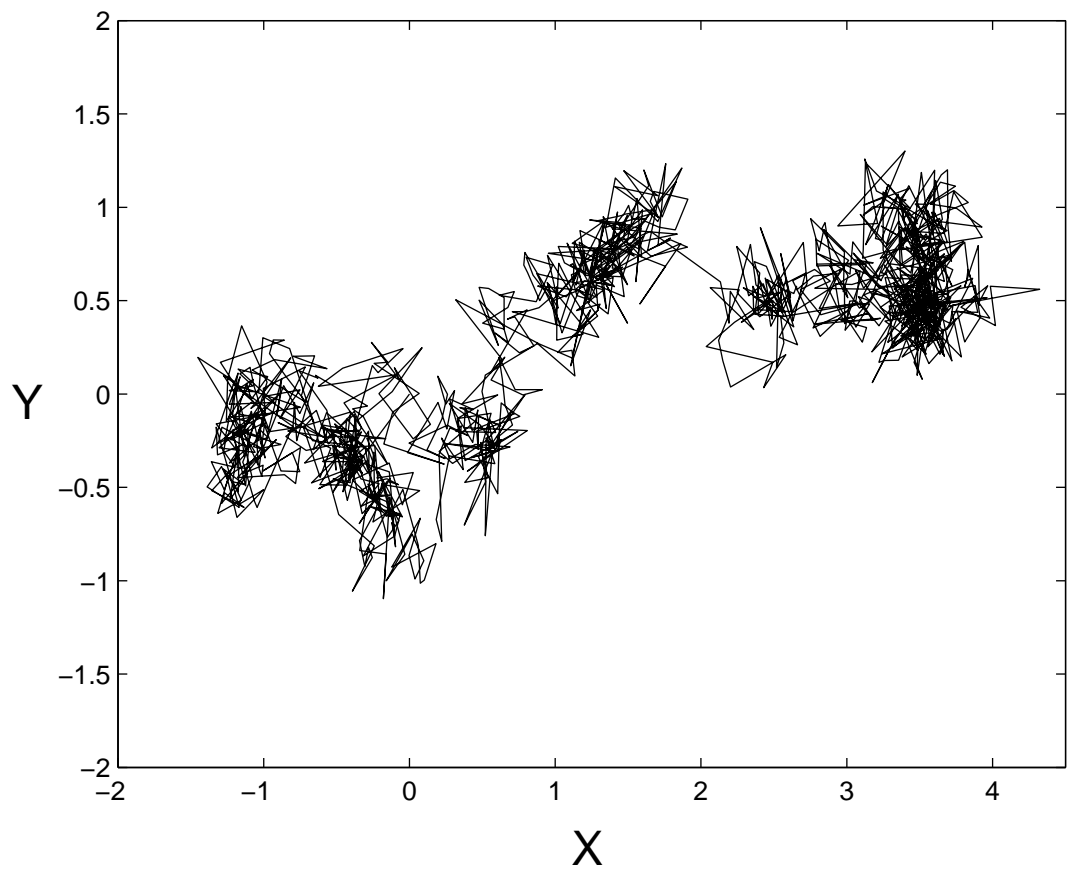
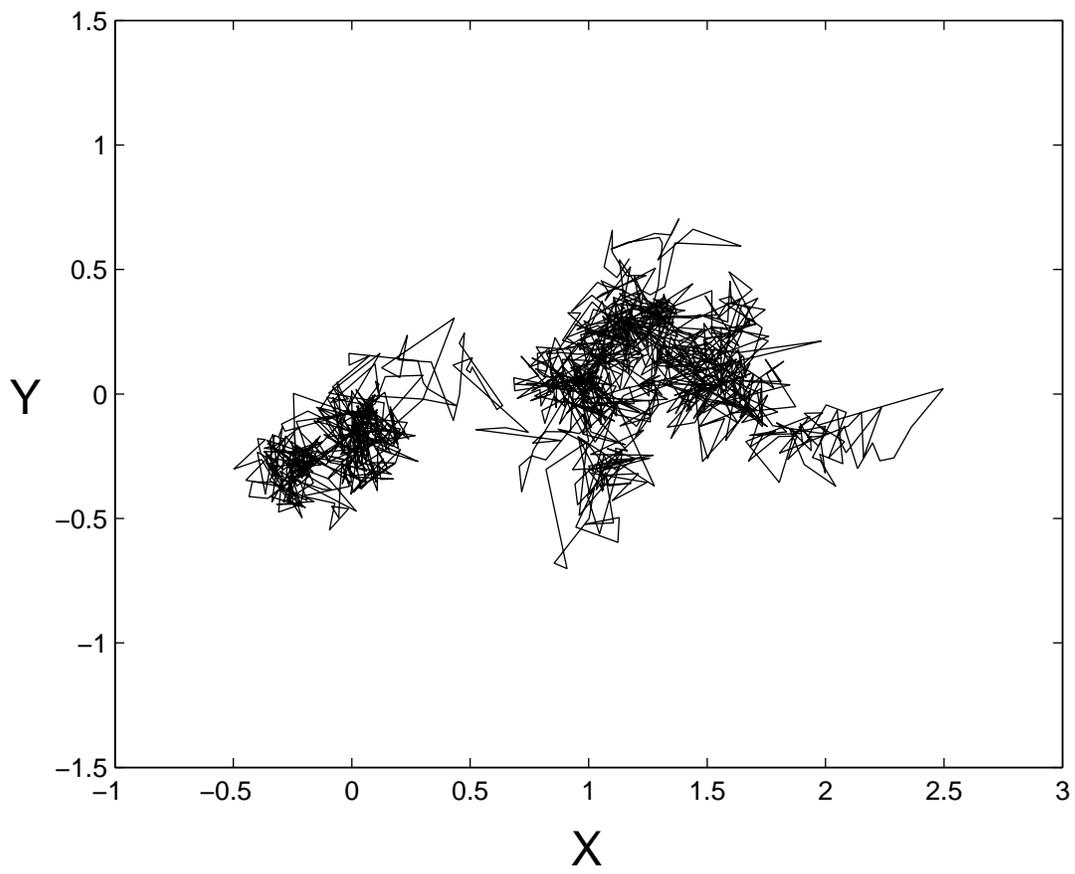
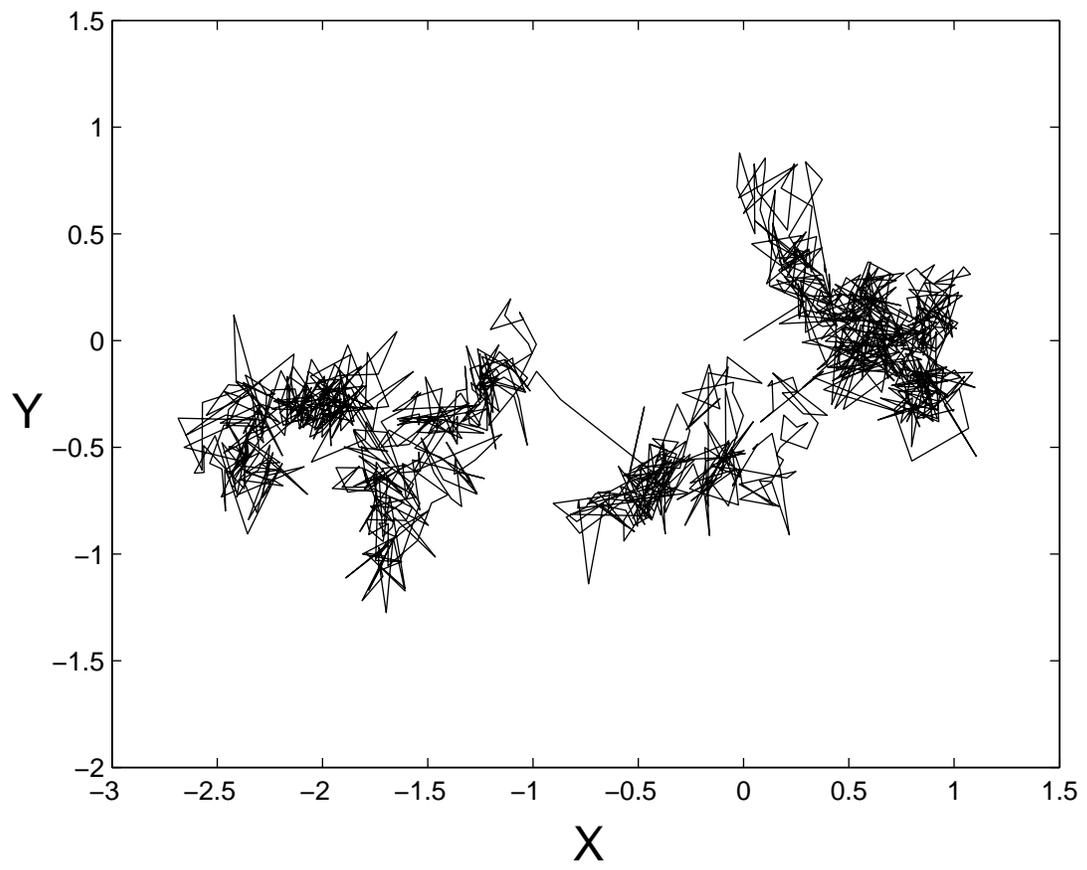
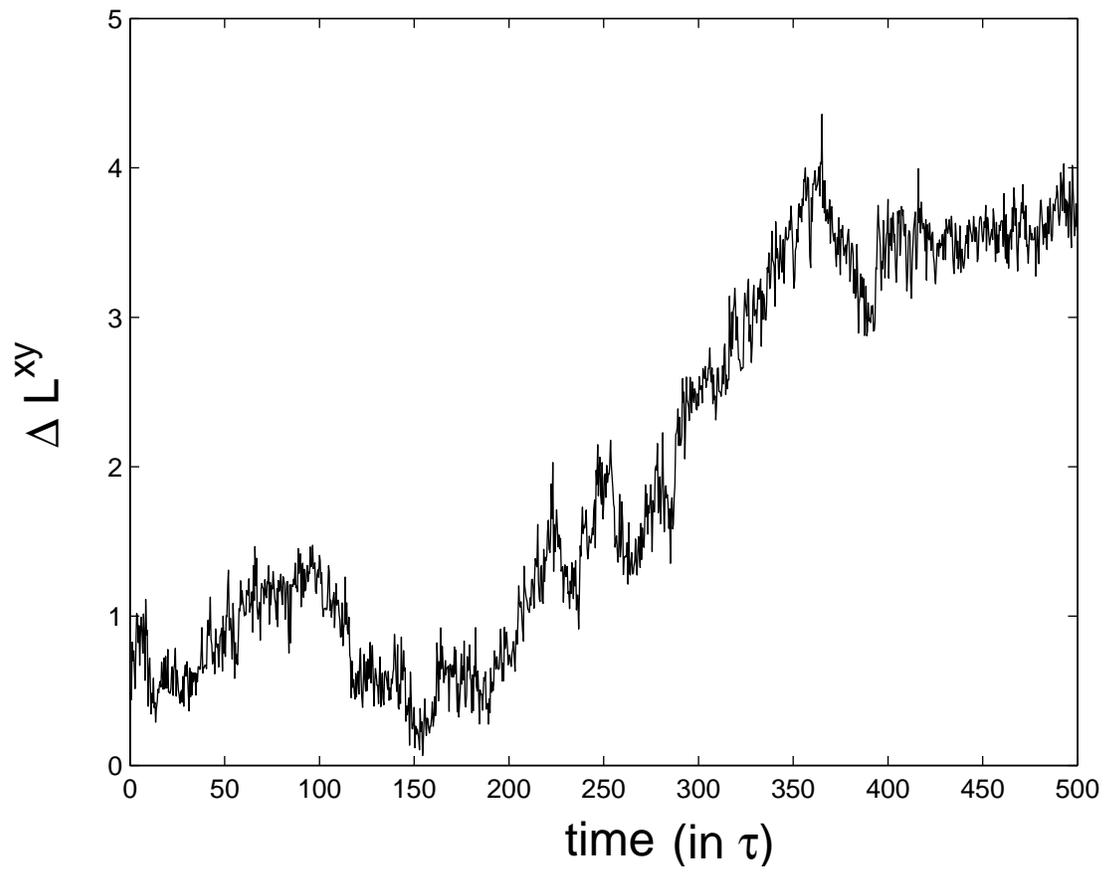
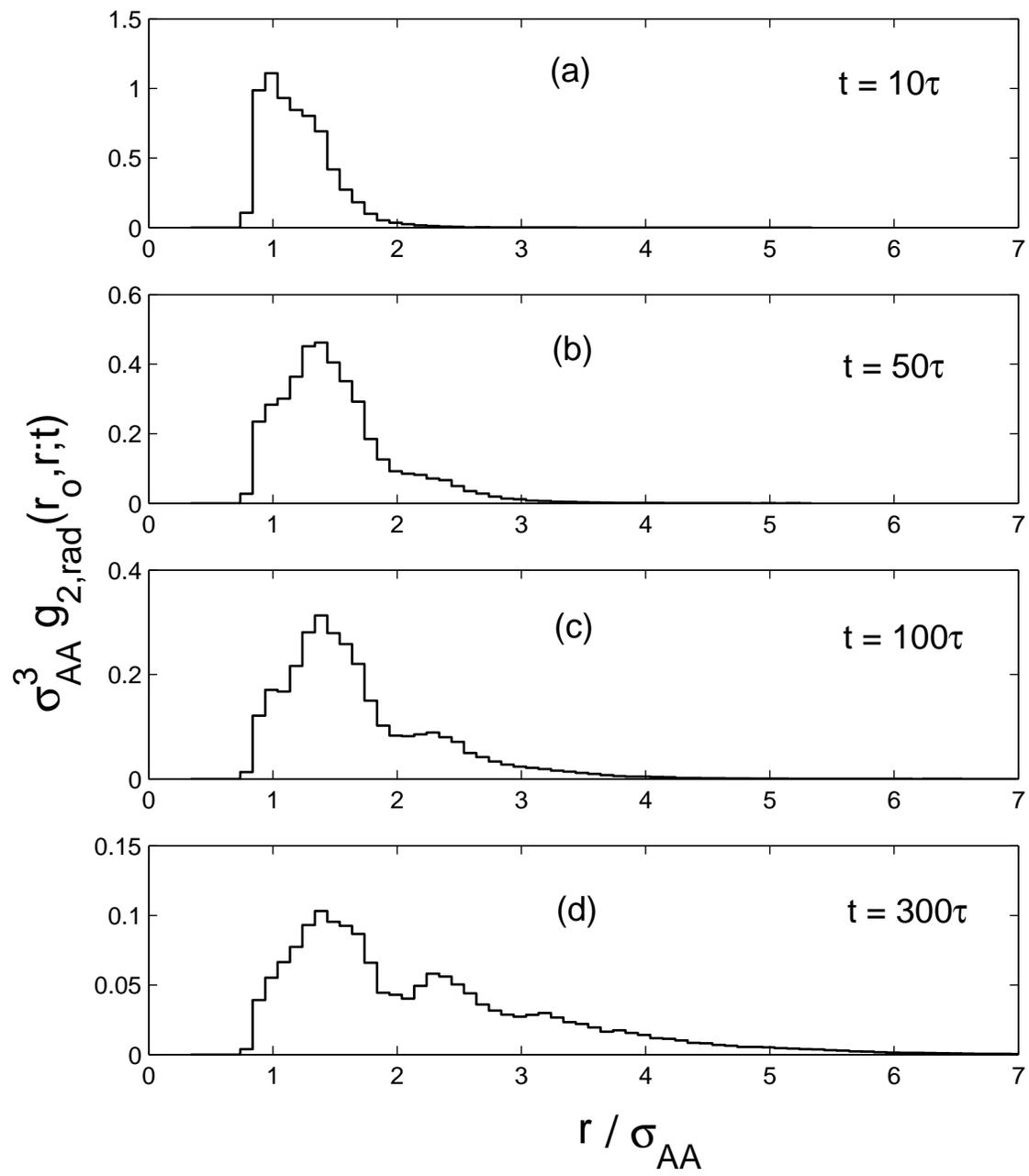
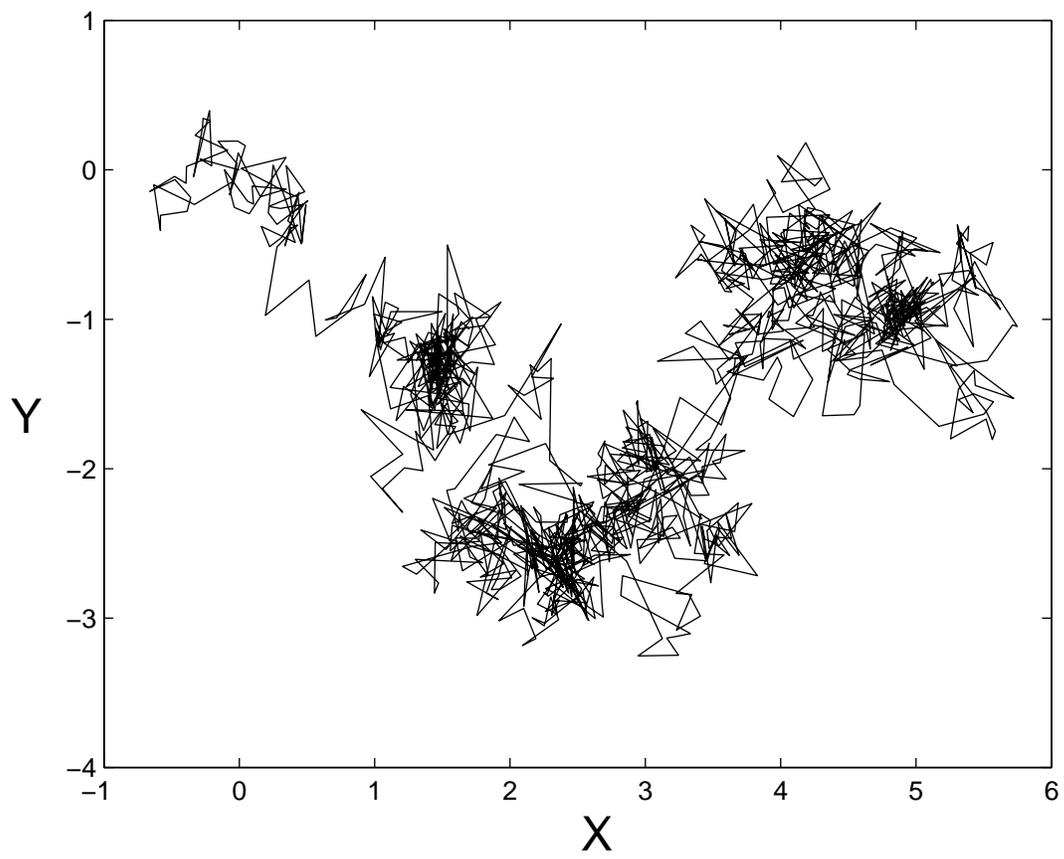
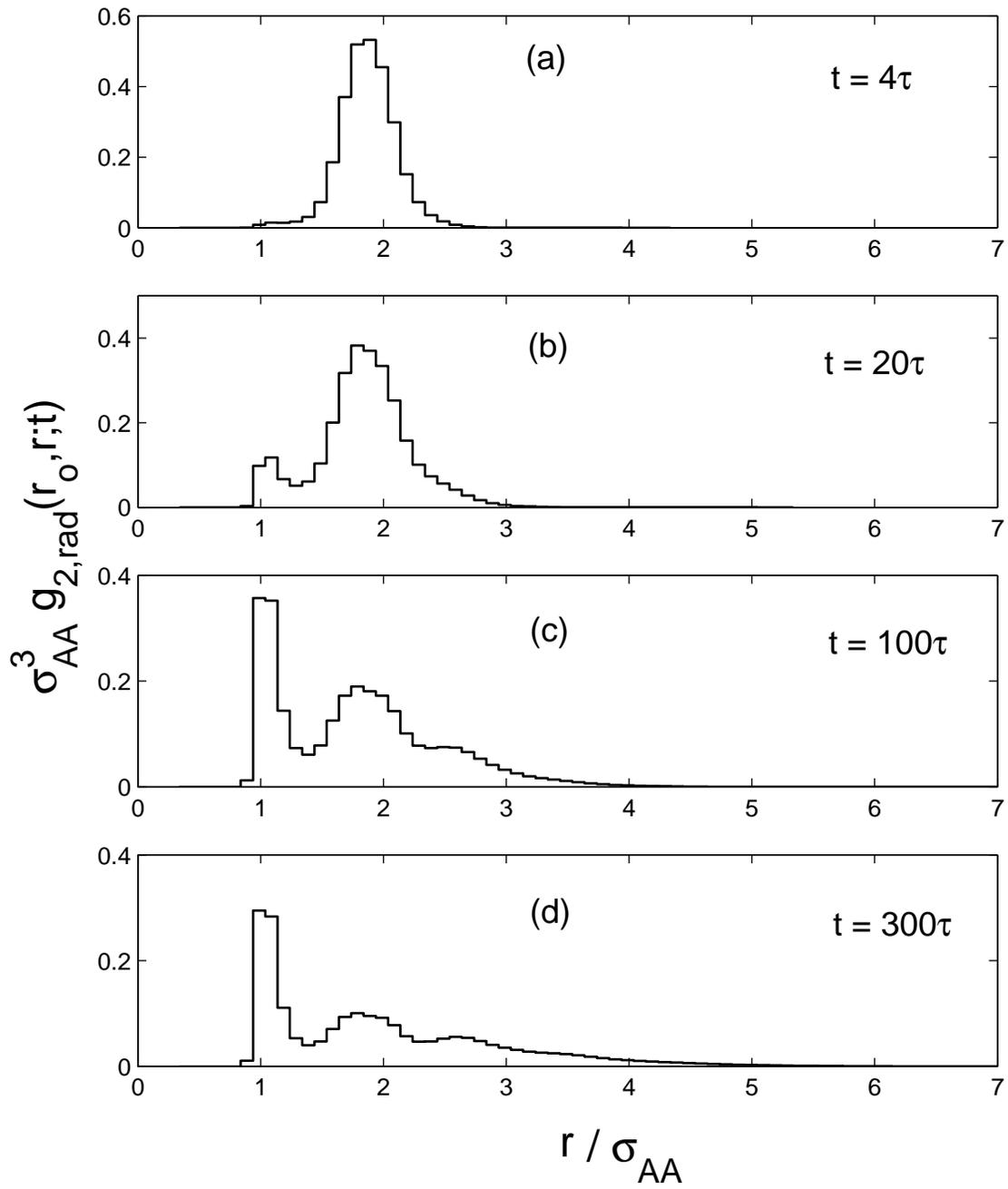
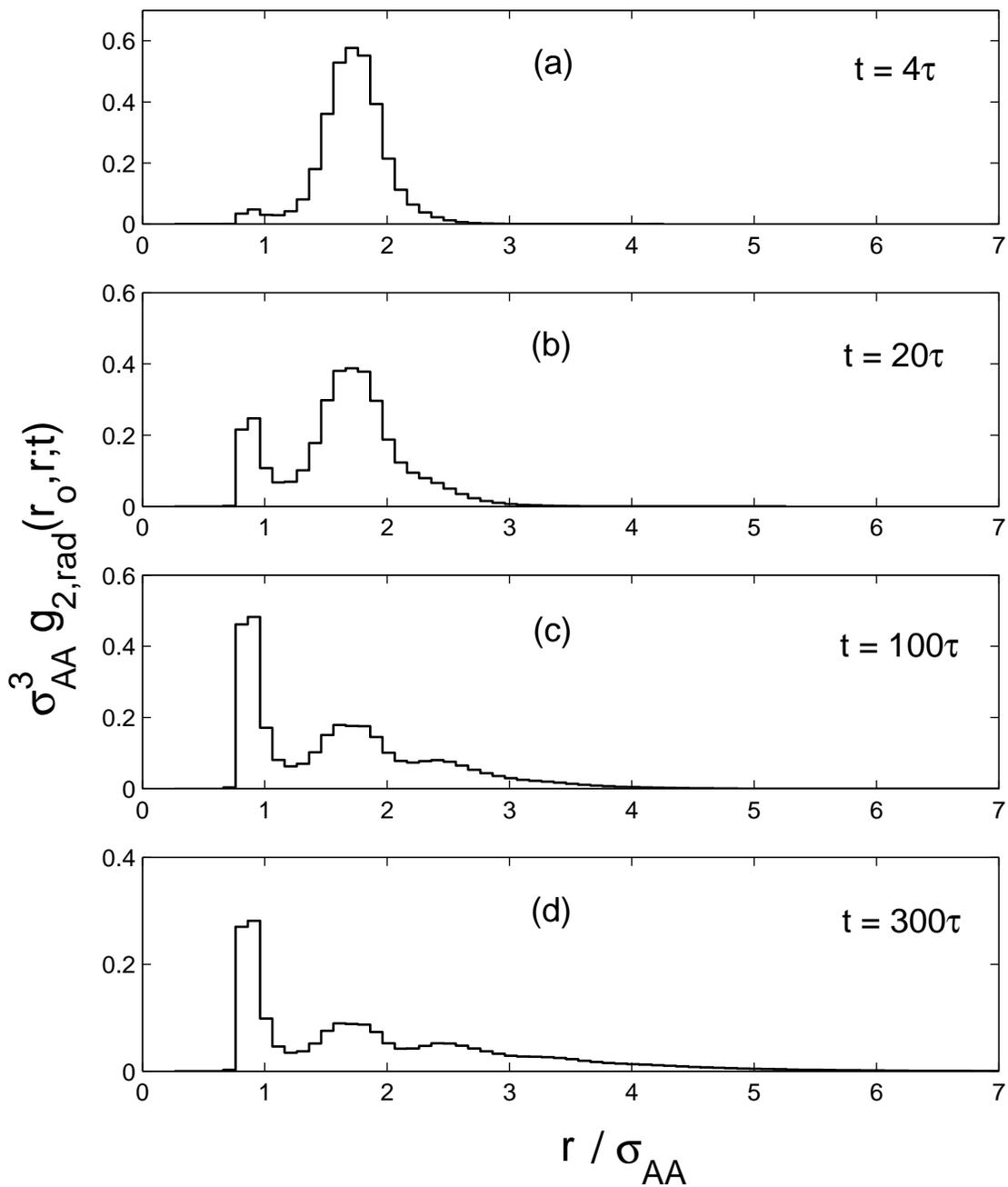
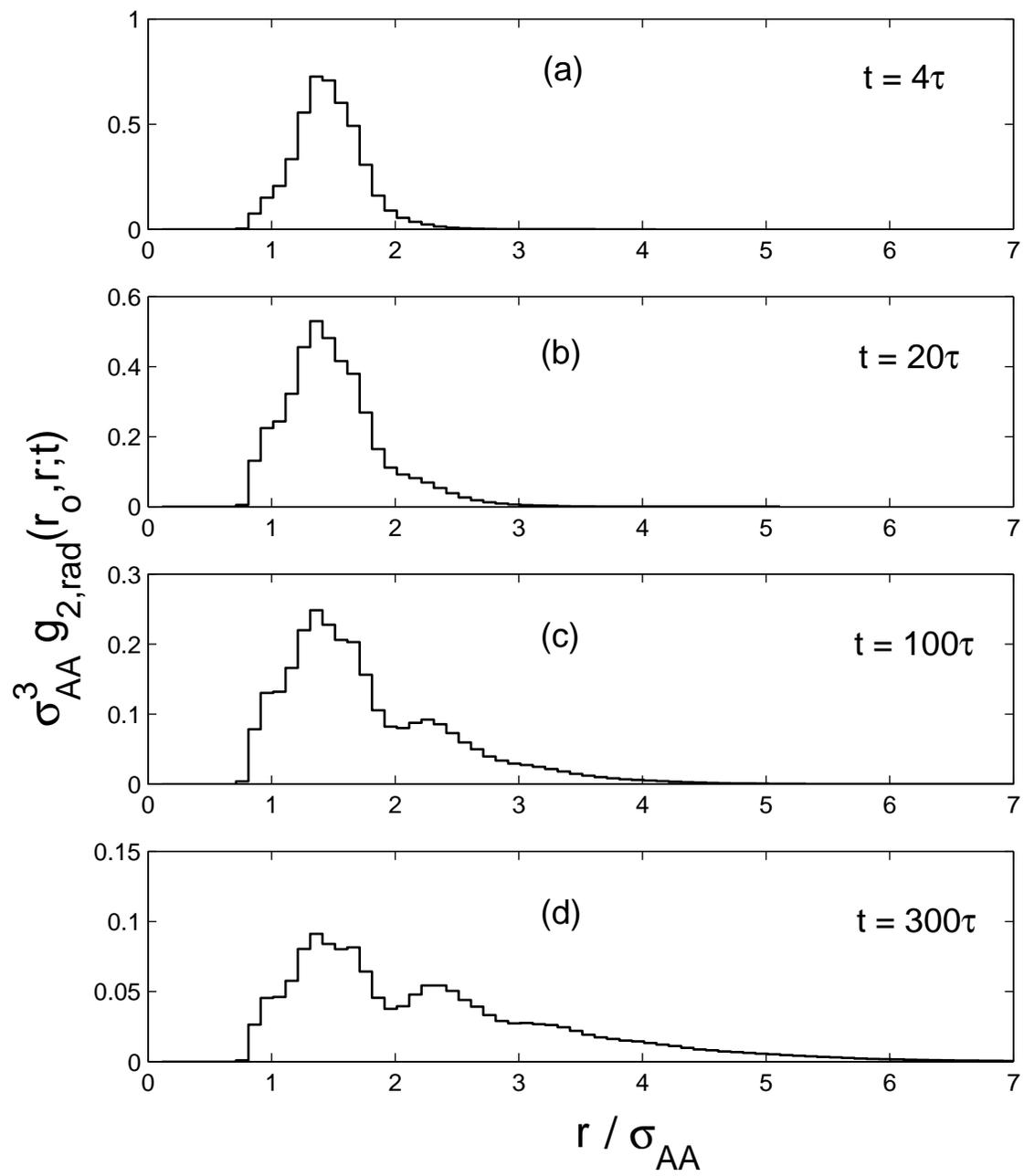
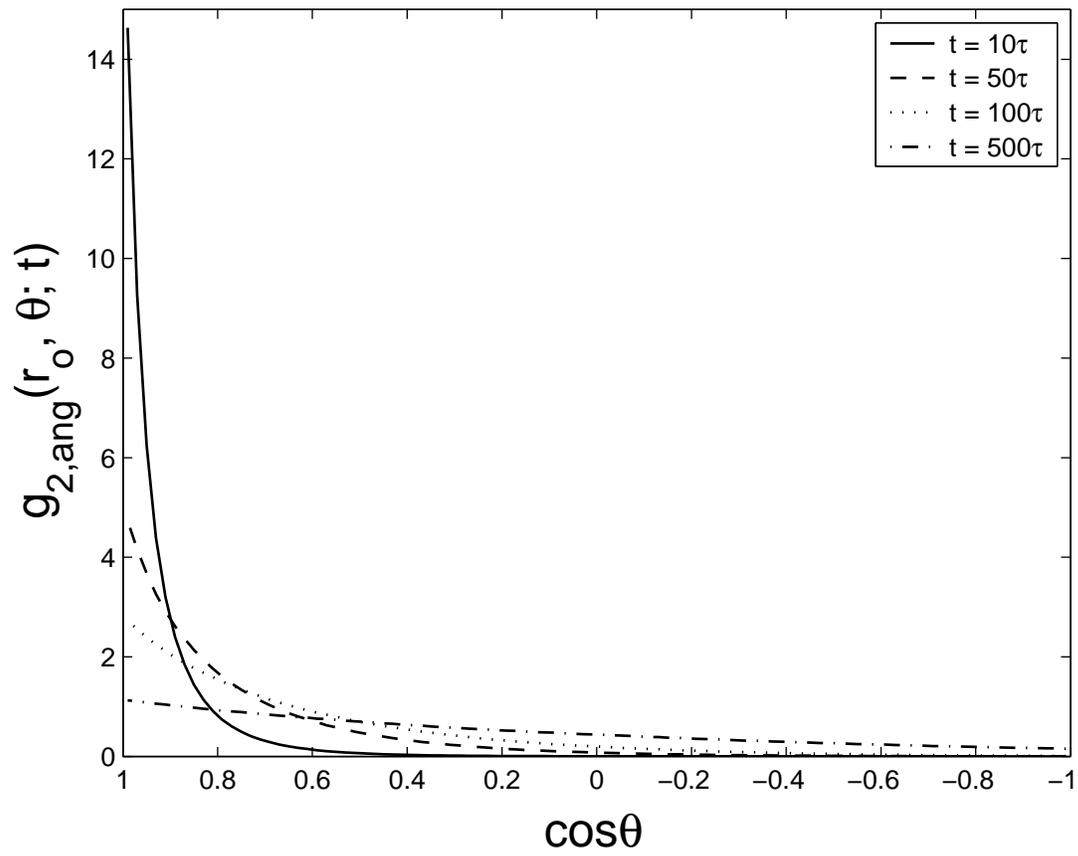
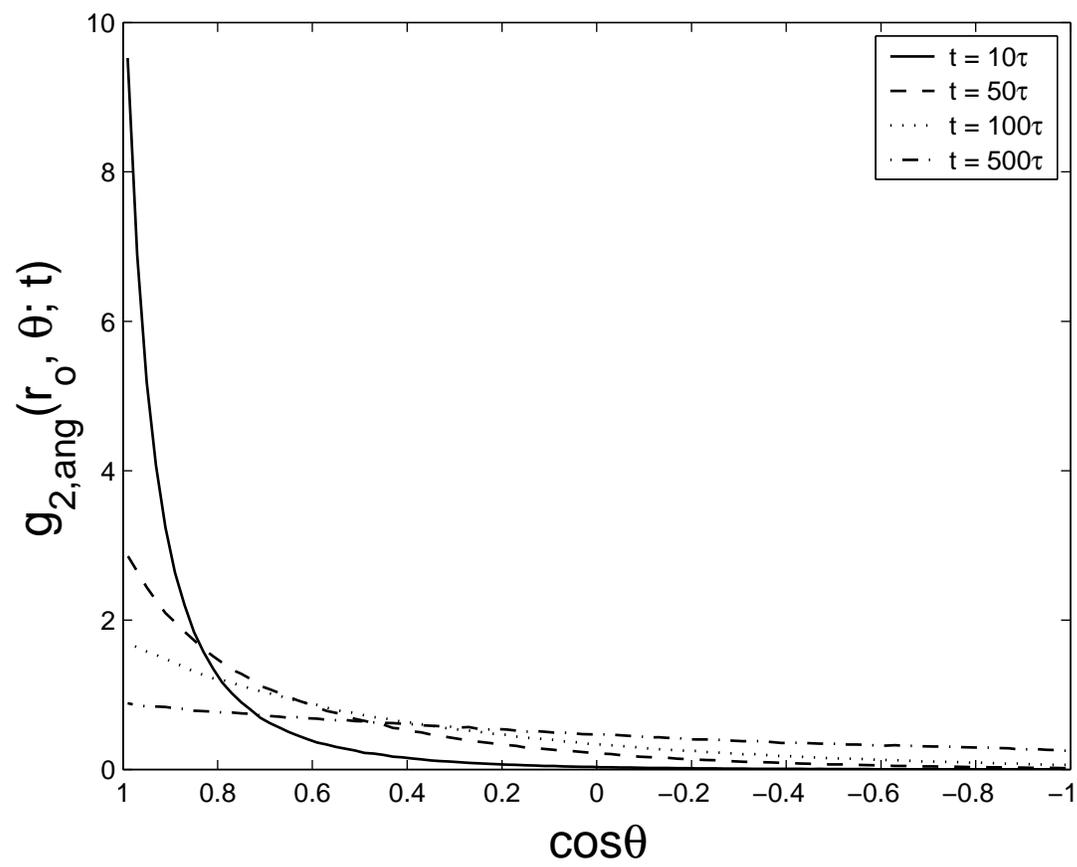
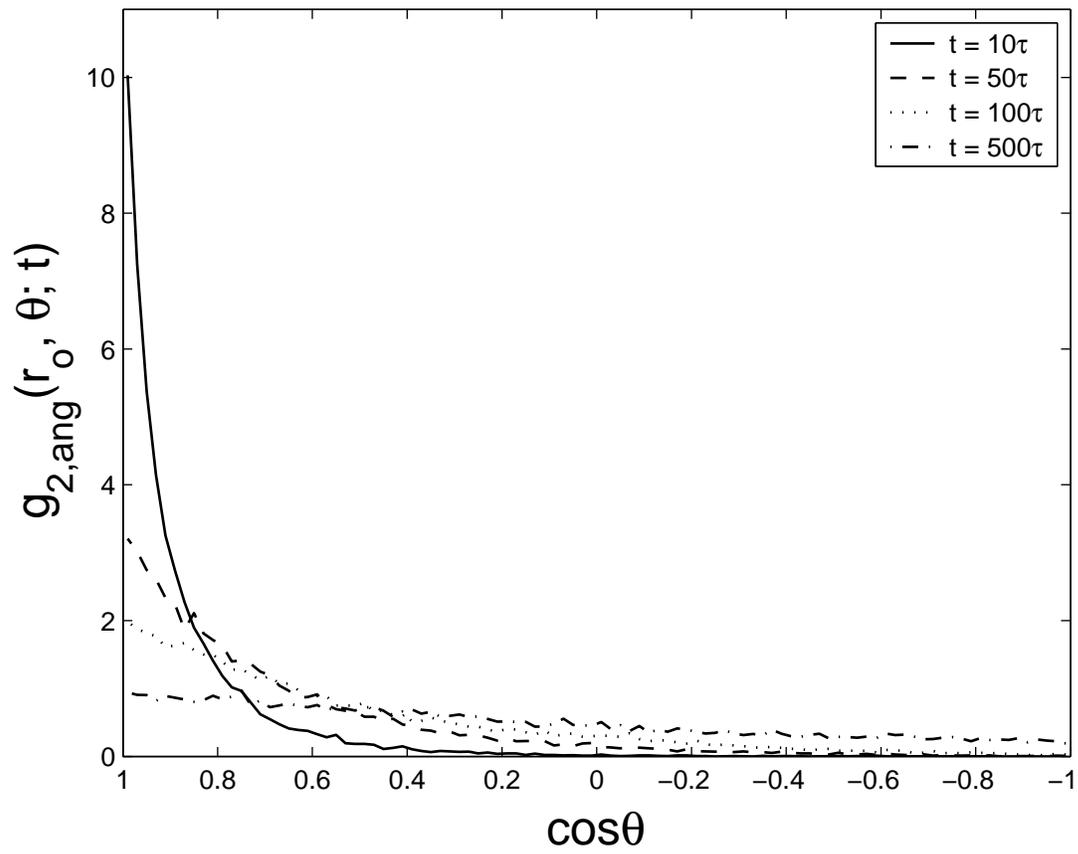
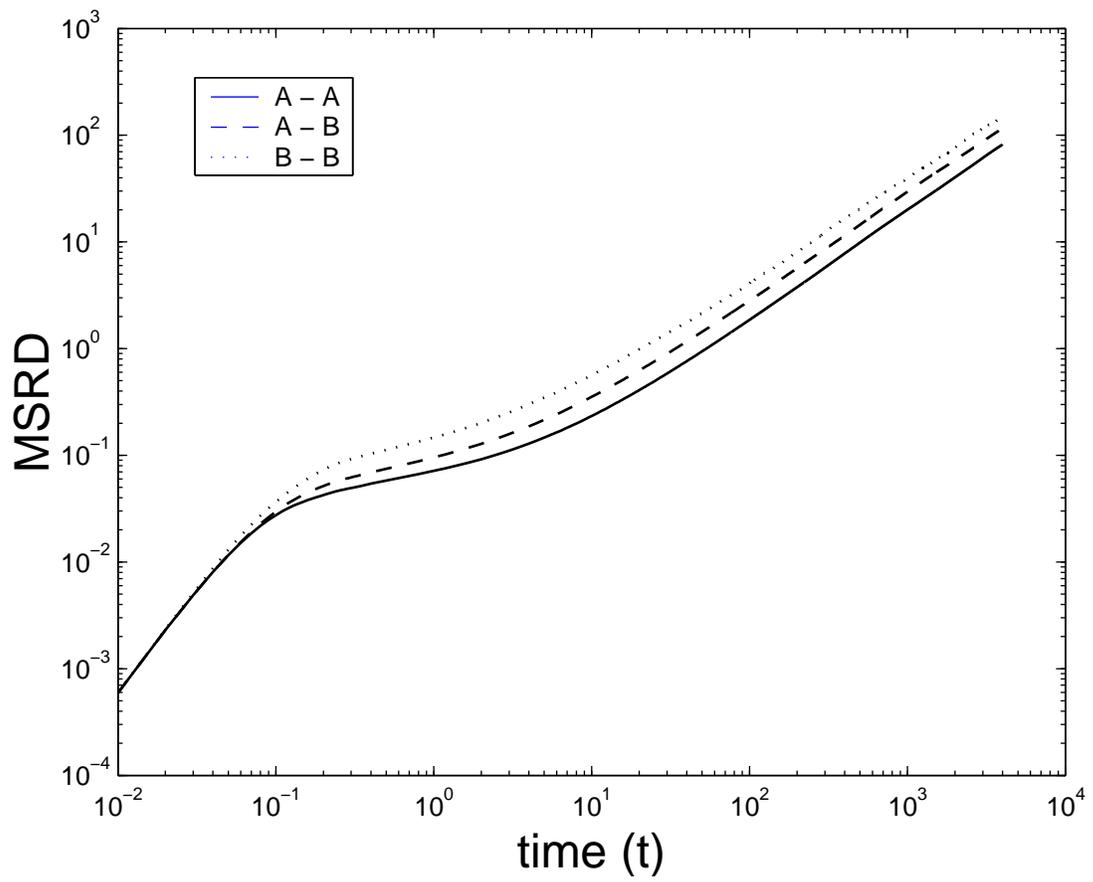
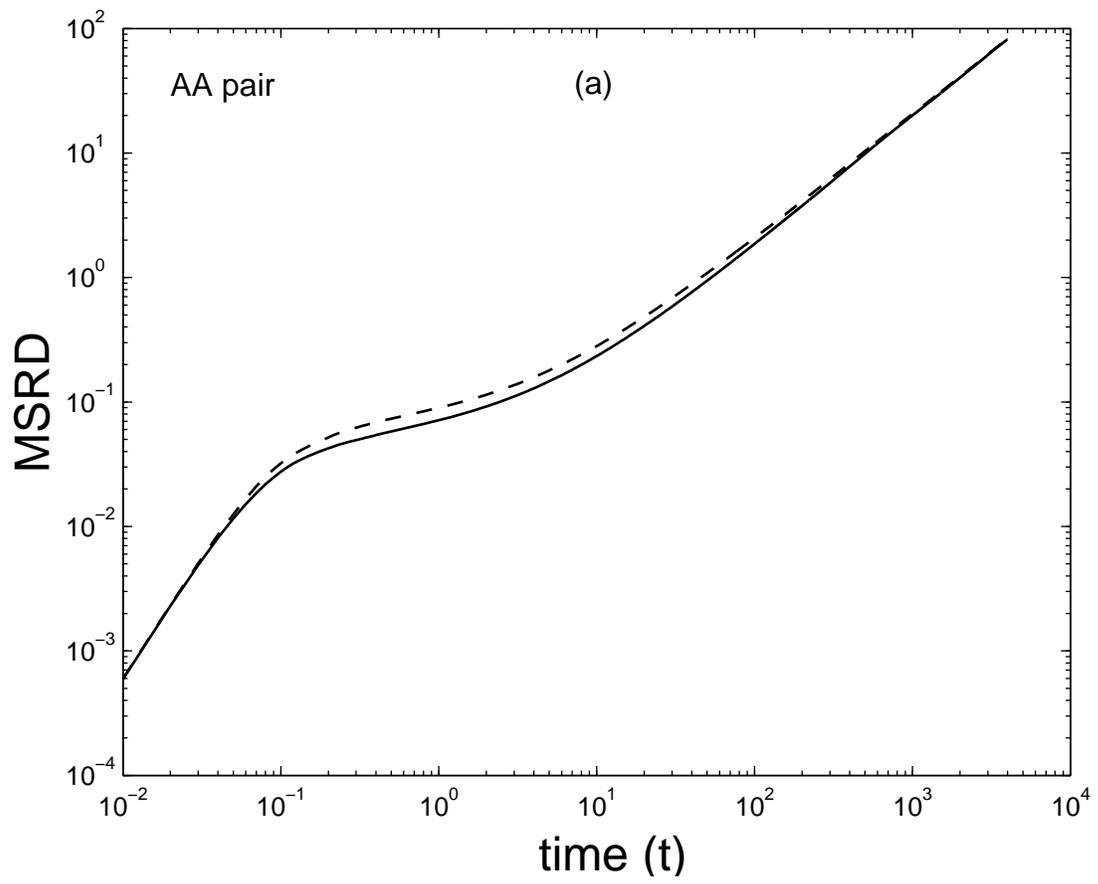
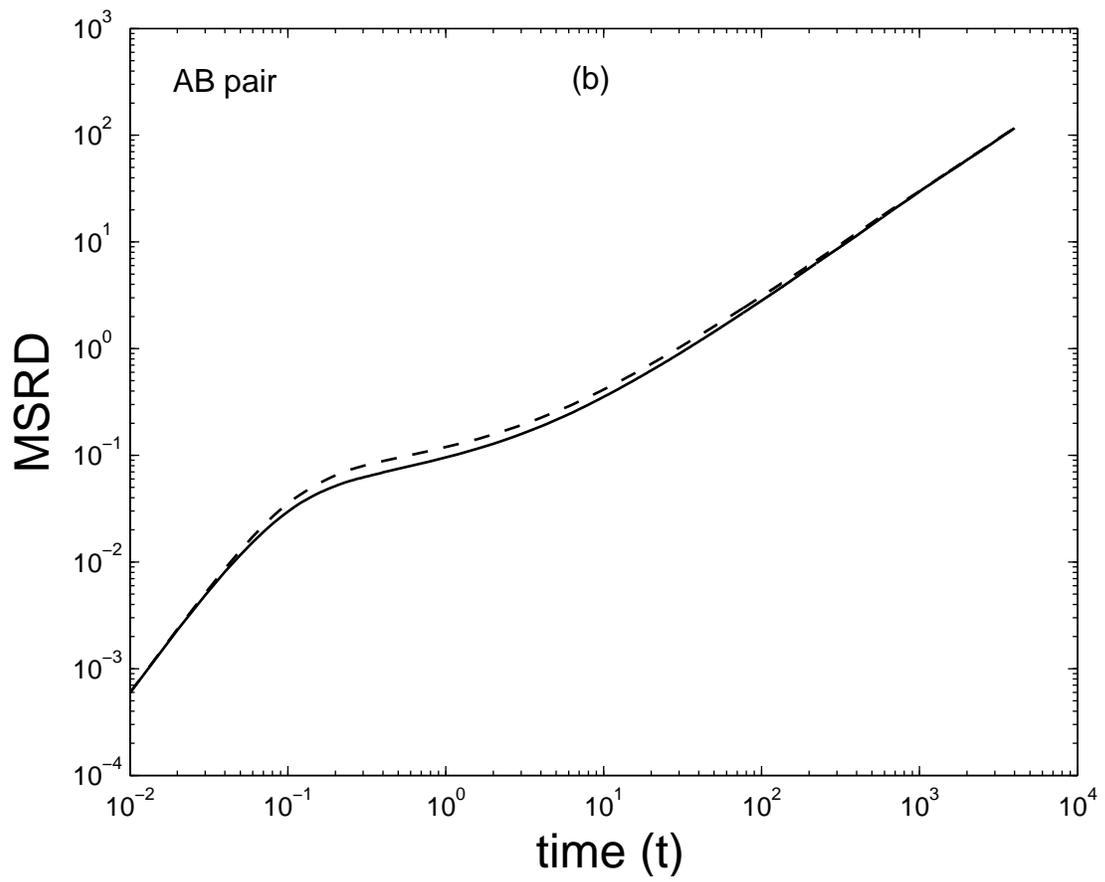
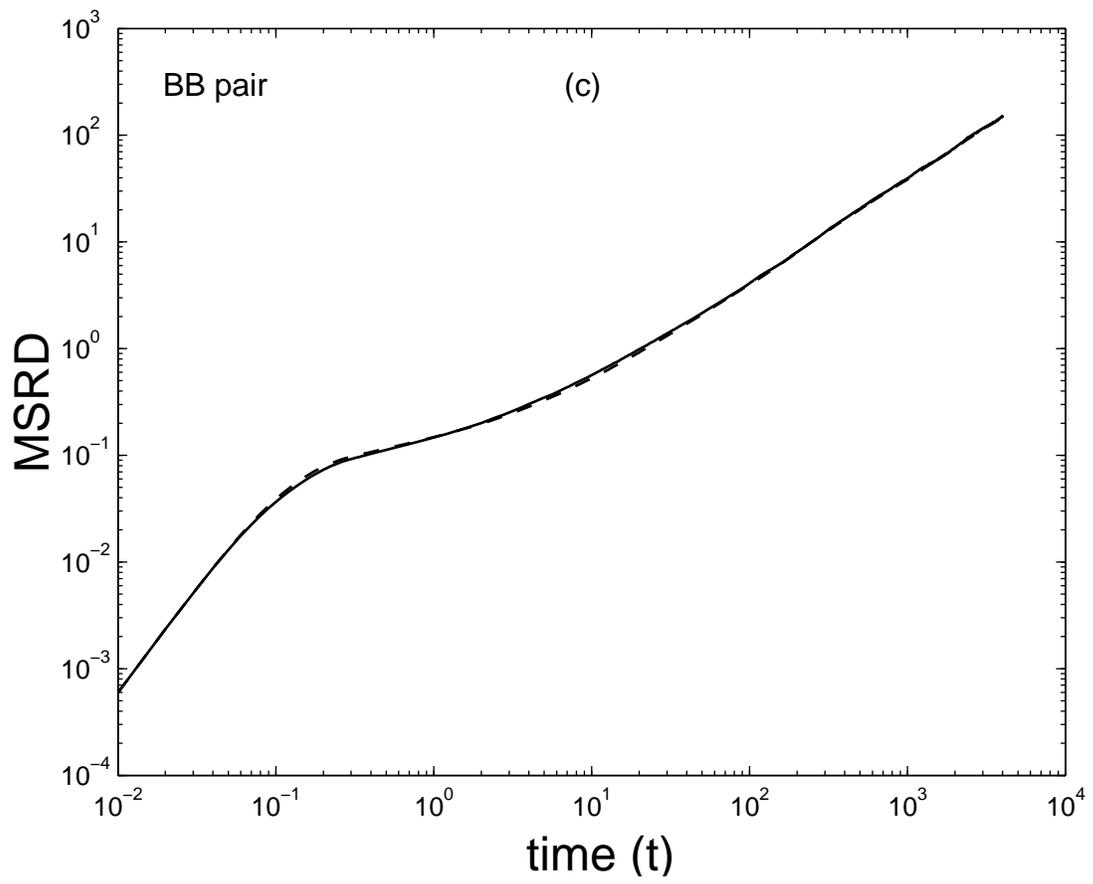
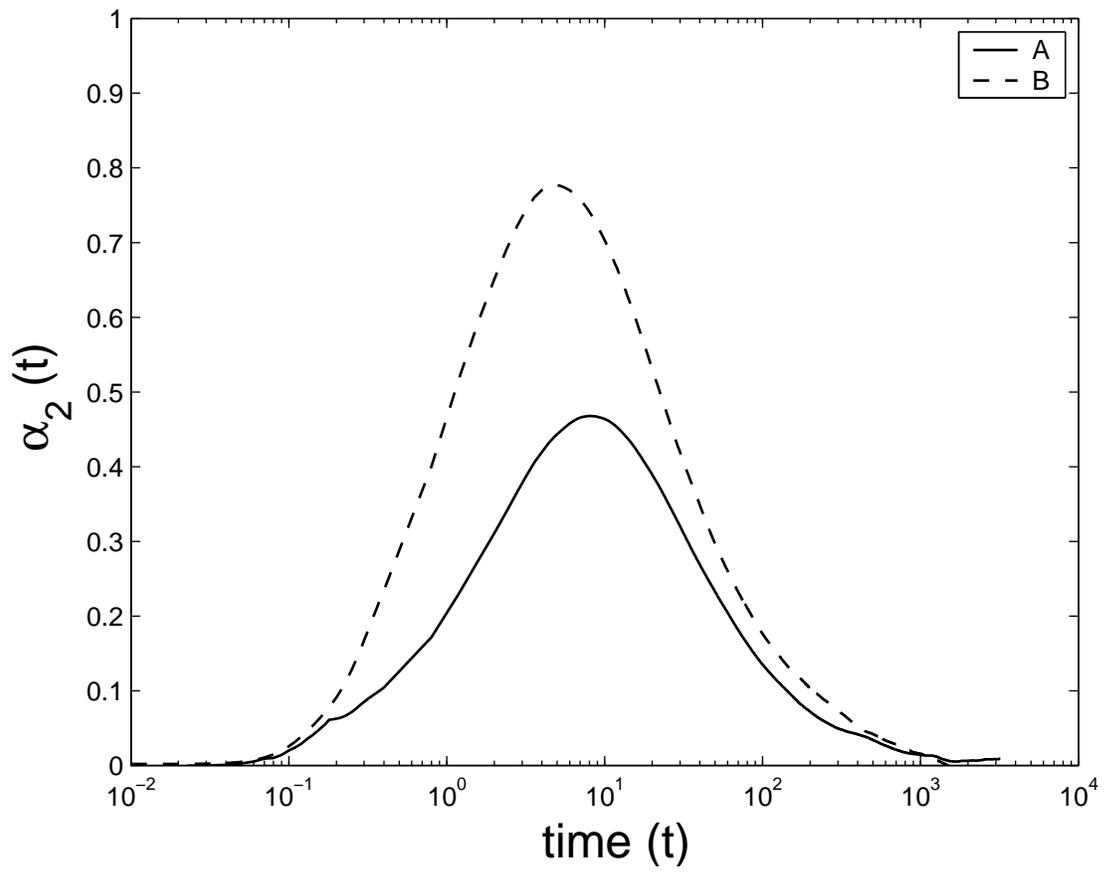
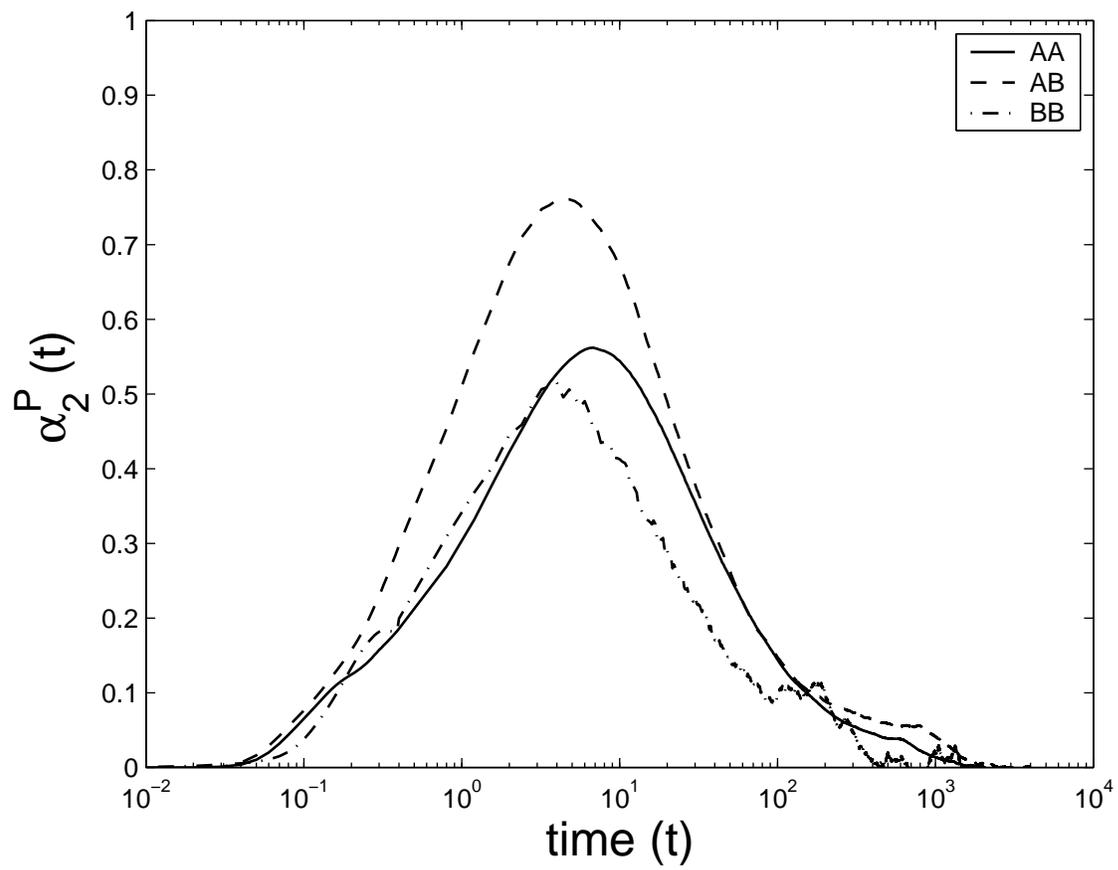
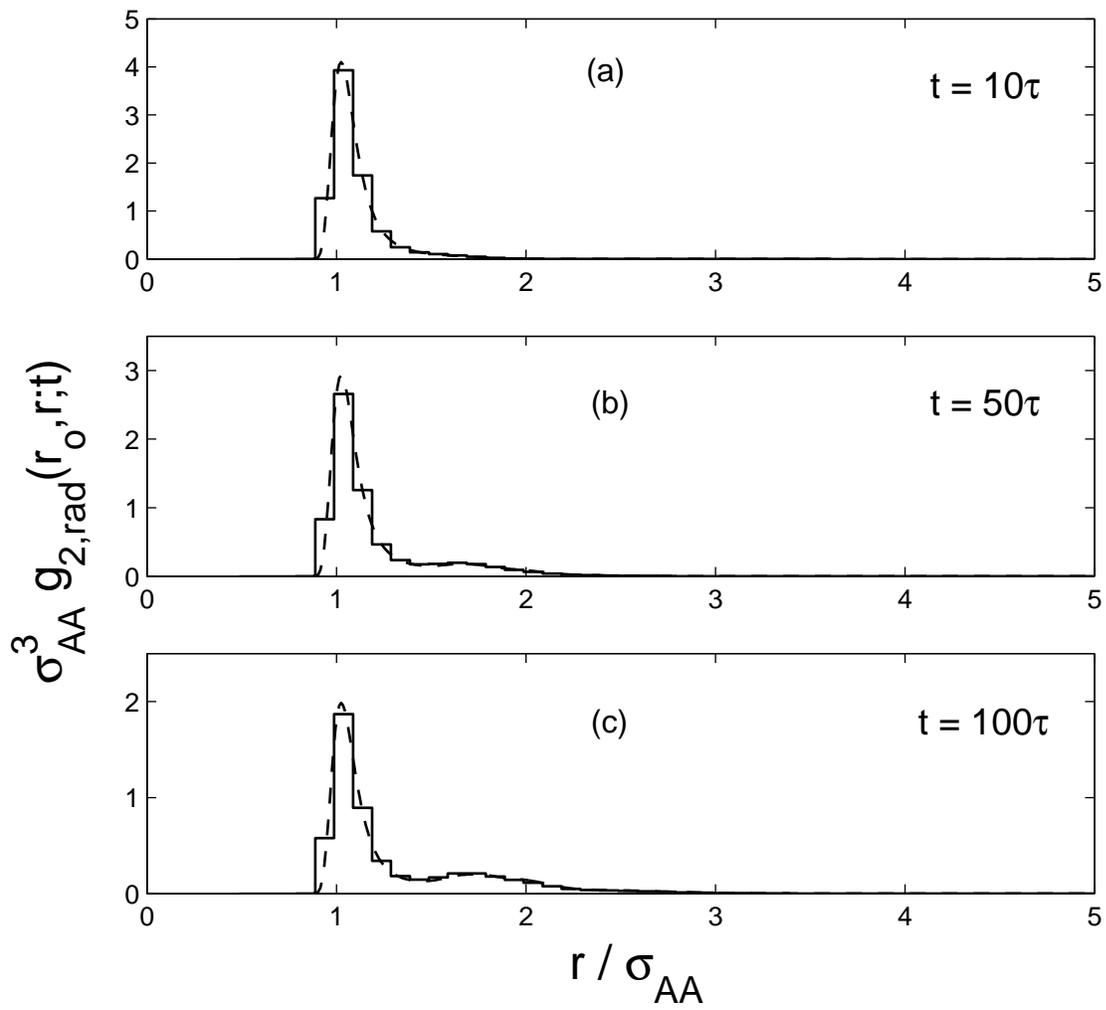
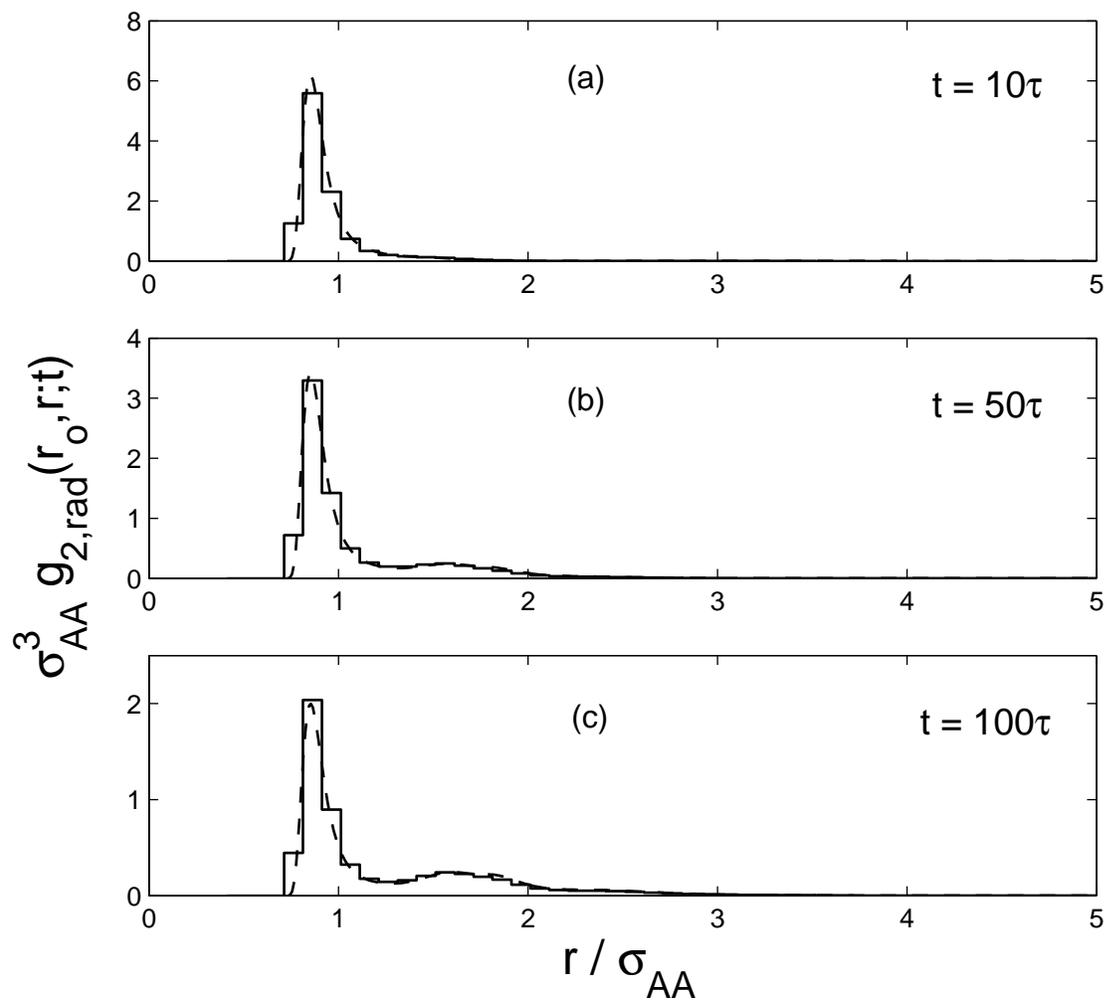
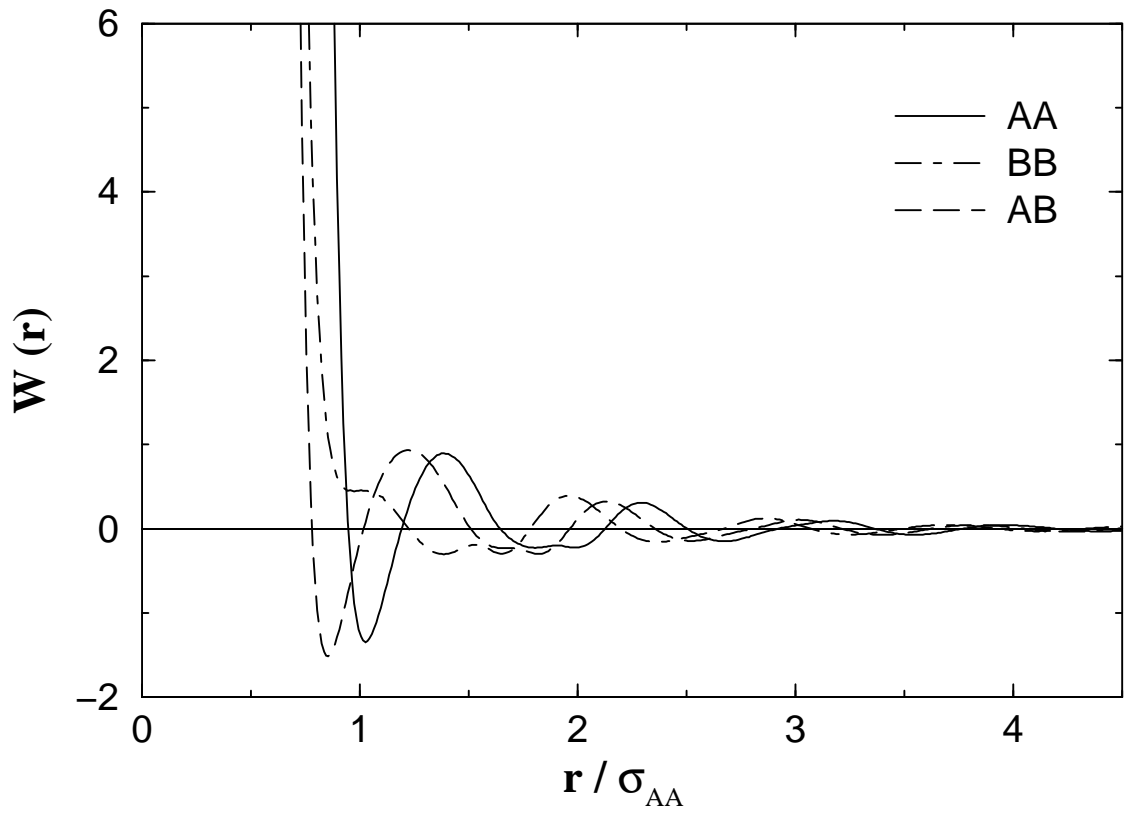
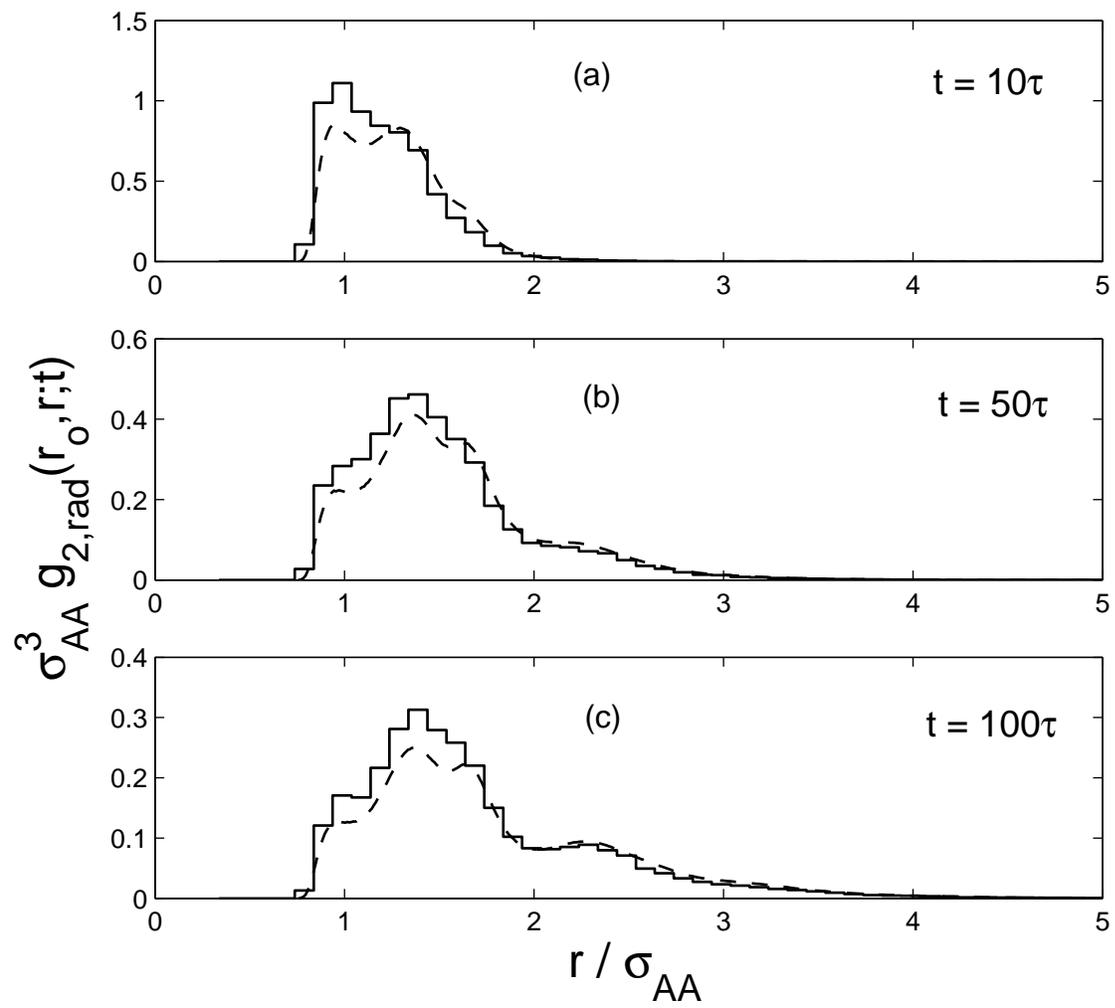
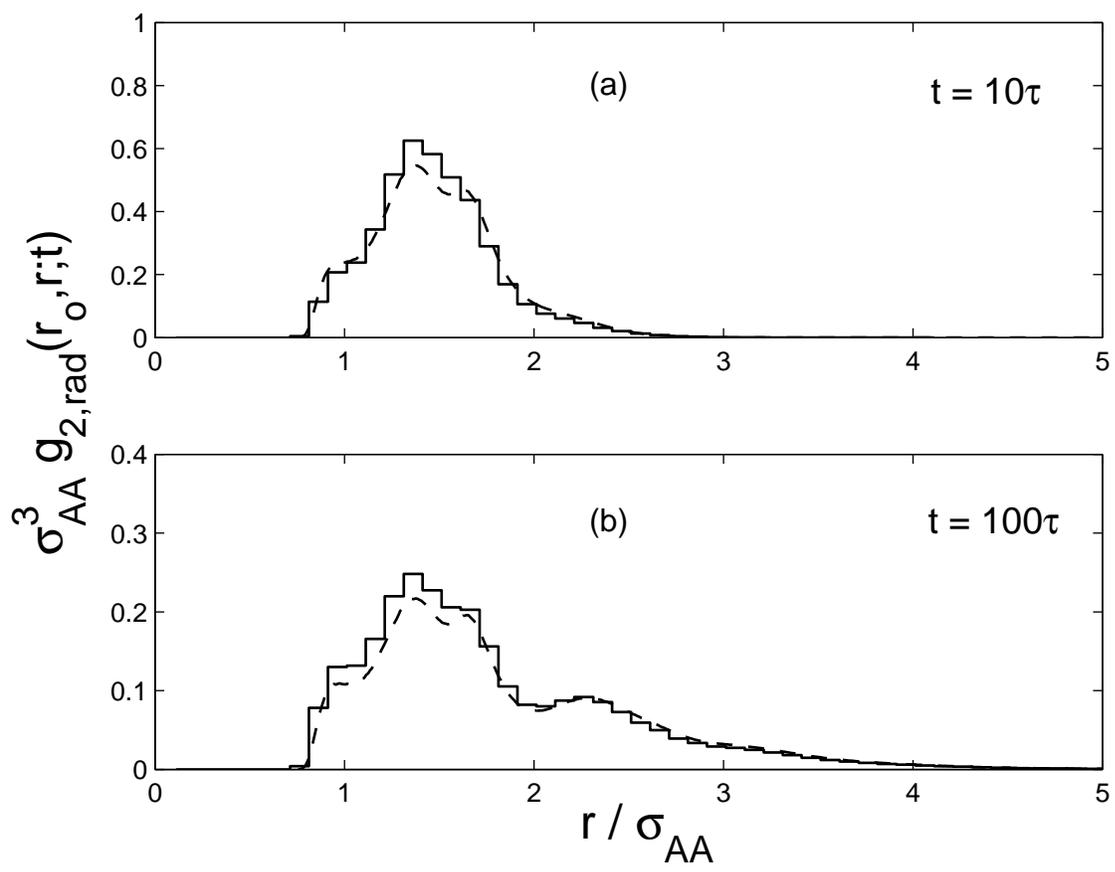
\end{document}